\documentclass{emulateapj}
\usepackage{apjfonts}
\def\mum{\mu\mathrm{m}}
\def\FIivAfe{\mathrm{F}^{\,\mathrm{afe}}_{8}}

\def\LIivAfe{\mathrm{L}^{\,\mathrm{afe}}_{\,8}}
\def\LIrsIivAfe{\mathrm{L}^{\,\mathrm{afe}}_{\,8\,\mathrm{(S)}}}
\def\FIrsIivAfe{\mathrm{F}^{\,\mathrm{afe}}_{8\,\mathrm{(S)}}}
\def\FIrsMirAfe{\mathrm{F}^{\,\mathrm{afe}}_{\mathrm{mir}\,\mathrm{(S)}}}
\def\fIrsIivAfe{\mathrm{f}^{\,\mathrm{afe}}_{\,8\,\mathrm{(S)}}}

\def\fIrsIivCnt{\mathrm{f}^{\,\mathrm{cnt}}_{\,8\,\mathrm{(S)}}}
\def\fIrsIivAmo{\mathrm{f}^{\,\mathrm{nl}}_{\,8\,\mathrm{(S)}}}
\def\fIivNs{\mathrm{f}^{\,\mathrm{ns}}_{\,8}}
\def\fIivAfe{\mathrm{f}^{\,\mathrm{afe}}_{\,8}}
\def\fIivCnt{\mathrm{f}^{\,\mathrm{cnt}}_{\,8}}

\def\fIivStr{\mathrm{f}^{\,\mathrm{str}}_{\,8}}
\def\fIiiiNs{\mathrm{f}^{\,\mathrm{ns}}_{\,5.8}}
\def\fIiiiStr{\mathrm{f}^{\,\mathrm{str}}_{\,5.8}}
\def\fIiiNs{\mathrm{f}^{\,\mathrm{ns}}_{\,4.5}}
\def\fIiiStr{\mathrm{f}^{\,\mathrm{str}}_{\,4.5}}
\def\fIiStr{\mathrm{f}^{\,\mathrm{str}}_{\,3.6}}

\def\fMiStr{\mathrm{f}^{\,\mathrm{str}}_{\,24}}
\def\fMiNs{\mathrm{f}^{\,\mathrm{ns}}_{\,24}}
\def\fXNs{\mathrm{f}^{\,\mathrm{ns}}_{\,\mathrm{X}}}
\def\fXStr{\mathrm{f}^{\,\mathrm{str}}_{\,\mathrm{X}}}
\def\fX{\mathrm{f}_{\,\mathrm{X}}}
\def\fIiv{\mathrm{f}_{\,8}}
\def\FIiv{\mathrm{F}_{8}}
\def\LIiv{\mathrm{L}_{\,8}}
\def\fIiii{\mathrm{f}_{\,5.8}}
\def\fIii{\mathrm{f}_{\,4.5}}
\def\fIi{\mathrm{f}_{\,3.6}}
\def\fMi{\mathrm{f}_{\,24}}
\def\fIrsMi{\mathrm{f}_{\,24\,\mathrm{(S)}}}
\def\fMii{\mathrm{f}_{\,70}}
\def\fMiii{\mathrm{f}_{\,160}}
\def\FTir{\mathrm{F}_{\mathrm{TIR}}}
\def\LTir{\mathrm{L}_{\mathrm{TIR}}}
\def\Lsun{\mathrm{L}_{\sun}}
\def\so{1}
\def\sotxt{Steward Observatory, University of Arizona, Tucson, AZ 85721 USA}
\def\indiana{2}
\def\indianatxt{Department of Astronomy, Indiana University, Bloomington, IN 47405 USA}
\def\wyoming{3}
\def\wyomingtxt{Department of Physics and Astronomy, University of Wyoming, Laramie, WY 82071 USA}
\def\toledo{4}
\def\toledotxt{Ritter Astrophysical Research Center, University of Toledo, Toledo, OH 43603}
\def\stsci{5}
\def\stscitxt{Space Telescope Science Institute, Baltimore, MD 21218 USA}
\def\ipac{6}
\def\ipactxt{Infrared Processing and Analysis Center, California Institute of Technology, Pasadena, CA 91125 USA}
\def\carnegie{7}
\def\carnegietxt{Carnegie Observatories, Pasadena, CA 91101 USA}
\def\cambridge{8}
\def\cambridgetxt{Institute of Astronomy, University of Cambridge, Cambridge CB3 0HA, UK}
\def\minnesota{9}
\def\minnesotatxt{Astronomy Department, University of Minnesota, Minneapolis, MN 55455 USA}
\def\washington{10}
\def\washingtontxt{Department of Astronomy, University of Washington, Seattle, WA 98195 USA}
\def\umass{11}
\def\umasstxt{Astronomy Department, University of Massachusetts, Amherst, MA 01003 USA}
\def\gemini{12}
\def\geminitxt{Gemini Observatory, Southern Operations Center, Casilla 603, La Serena, Chile}

\slugcomment{}

\shorttitle{AFE in LVL}
\shortauthors{Marble et al.}
\journalinfo{Accepted to the Astrophysical Journal on April 7, 2010}

\begin{document}

\title{An Aromatic Inventory of the Local Volume}

\author
{
  A.~R.~Marble\altaffilmark{\so}, 
  C.~W.~Engelbracht\altaffilmark{\so},  
  L.~van~Zee\altaffilmark{\indiana},
  D.~A.~Dale\altaffilmark{\wyoming},
  J.~D.~T.~Smith\altaffilmark{\toledo},
  K.~D.~Gordon\altaffilmark{\stsci},
  Y.~Wu\altaffilmark{\ipac},
  J.~C.~Lee\altaffilmark{\carnegie},
  R.~C.~Kennicutt\altaffilmark{\cambridge},
  E.~D.~Skillman\altaffilmark{\minnesota},
  L.~C.~Johnson\altaffilmark{\washington},
  M.~Block\altaffilmark{\so},
  D.~Calzetti\altaffilmark{\umass},
  S.~A.~Cohen\altaffilmark{\wyoming},
  H.~Lee\altaffilmark{\gemini},
  M.~D.~Schuster\altaffilmark{\wyoming}
}

\altaffiltext{\so}{\sotxt}
\altaffiltext{\indiana}{\indianatxt}
\altaffiltext{\wyoming}{\wyomingtxt}
\altaffiltext{\toledo}{\toledotxt}
\altaffiltext{\stsci}{\stscitxt}
\altaffiltext{\ipac}{\ipactxt}
\altaffiltext{\carnegie}{\carnegietxt}
\altaffiltext{\cambridge}{\cambridgetxt}
\altaffiltext{\minnesota}{\minnesotatxt}
\altaffiltext{\washington}{\washingtontxt}
\altaffiltext{\umass}{\umasstxt}
\altaffiltext{\gemini}{\geminitxt}


\begin{abstract}
Using infrared photometry from the \emph{Spitzer Space Telescope}, we perform 
the first inventory of aromatic feature emission (AFE, but also commonly 
referred to as PAH emission) for a statistically complete sample of 
star-forming galaxies in the local volume.
The photometric methodology involved is calibrated and demonstrated to recover 
the aromatic fraction of the IRAC $8\mum$ flux with a standard deviation of 6\% 
for a training set
of 40 SINGS galaxies (ranging from stellar to dust dominated)
with both suitable mid-infrared \emph{Spitzer} IRS spectra 
and equivalent photometry.
A potential factor of two improvement could be realized with suitable $5.5\mum$
and $10\mum$
photometry, such as what may be provided in the future by \emph{JWST}. 
The resulting technique is then applied to mid-infrared photometry for 
the 258 galaxies from the Local Volume Legacy (LVL) survey,
a large sample dominated in number by low-luminosity dwarf 
galaxies for which obtaining comparable mid-infrared spectroscopy
is not feasible.  
We find the total LVL luminosity due to five strong aromatic features in the 
$8\mum$ complex to be 
$2.47\times10^{10}\,\Lsun$ with a mean volume density 
of $8.8\times10^6\,\Lsun\,\mathrm{Mpc}^{-3}$.  
Twenty-four of the LVL galaxies, corresponding to a luminosity cut at 
$\mathrm{M}_{\mathrm{B}} = -18.22$,
account for 90\% of the aromatic luminosity.
Using oxygen abundances compiled from the literature for 129 of the 258
LVL galaxies, we find a correlation between
metallicity and the aromatic to total infrared emission ratio but not the
aromatic to total $8\mum$ dust emission ratio.  A possible explanation is that
metallicity plays a role in the abundance of aromatic molecules relative to the
total dust content, but other factors such as star formation and/or the local 
radiation field affect the excitation of those molecules.
\end{abstract}




\section{Introduction}

It is well established that the infrared spectra of galaxies tend to be
dominated by emission from dust \citep[see, e.g., the review by][]{1979rie17araa477}.
This light is a significant fraction of the bolometric emission from galaxies,
making up roughly half the extragalactic background luminosity
\citep[e.g.,][]{2005lag43araa727}.  Investigation into the composition of the responsible dust
grains is hindered by the small number of known associated 
emission and absorption lines.  The strongest
features, by far, are a set of emission bands in the $3-17\,\mum$ range that are commonly observed
in galaxies.  In luminous
galaxies where the mid-infrared emission is dominated by star formation,
they appear to be ubiquitous \citep{1991roc248mnras606}
with little \citep{2003lu588apj199}, but not negligible
\citep{2007smi656apj770}, relative variation in feature strength.

These features have been referred to as the mid-infrared emission
bands, the unidentified infrared bands, aromatic emission features, and
variations thereof.  They have been tentatively identified as arising from
polycyclic aromatic hydrocarbons \citep[PAHs;][]{1984leg137aap5} and are often cited
as resulting from a mixture of such particles.  
Regardless of the exact composition, it is widely agreed that
the features in question arise from
various bending and stretching modes of aromatic molecules largely composed of
carbon and hydrogen; therefore, we refer to the resultant emission simply as 
aromatic emission or aromatic feature emission (AFE).

While present in nearly all star formation dominated galaxies,
aromatic features do exhibit some dependence on galaxy host properties.
It has been known for years that they can be weaker, or
absent altogether, in systems characterized by low metallicity and high star
formation intensity \citep{1991roc248mnras606,1999thu516apj783}.  \citet{2000mad44newar249}
suggested that metallicity might be an important parameter, and this
was shown to be true using the \emph{Spitzer Space Telescope} \citep{2004wer154apjs1} by 
\citet{2005eng628apj29} and others \citep{2006wu639apj157,2006jac646apj192,2008ros674apj814} and by detailed
examination of data from the \emph{Infrared Space Observatory}
\citep[ISO;][]{2006mad446aap877}.  
Further, the weakening of aromatic features in the low-metallicity outskirts
of the galaxy M101 was attributed by \citet{2008gor682apj336}
to the hardness of the radiation field, which was also suggested to be a factor by
the work of \citet{2006mad446aap877} and \citet{2008eng678apj804}.

\begin{figure*}[t] 
\begin{center}
\includegraphics[width=7.1in]{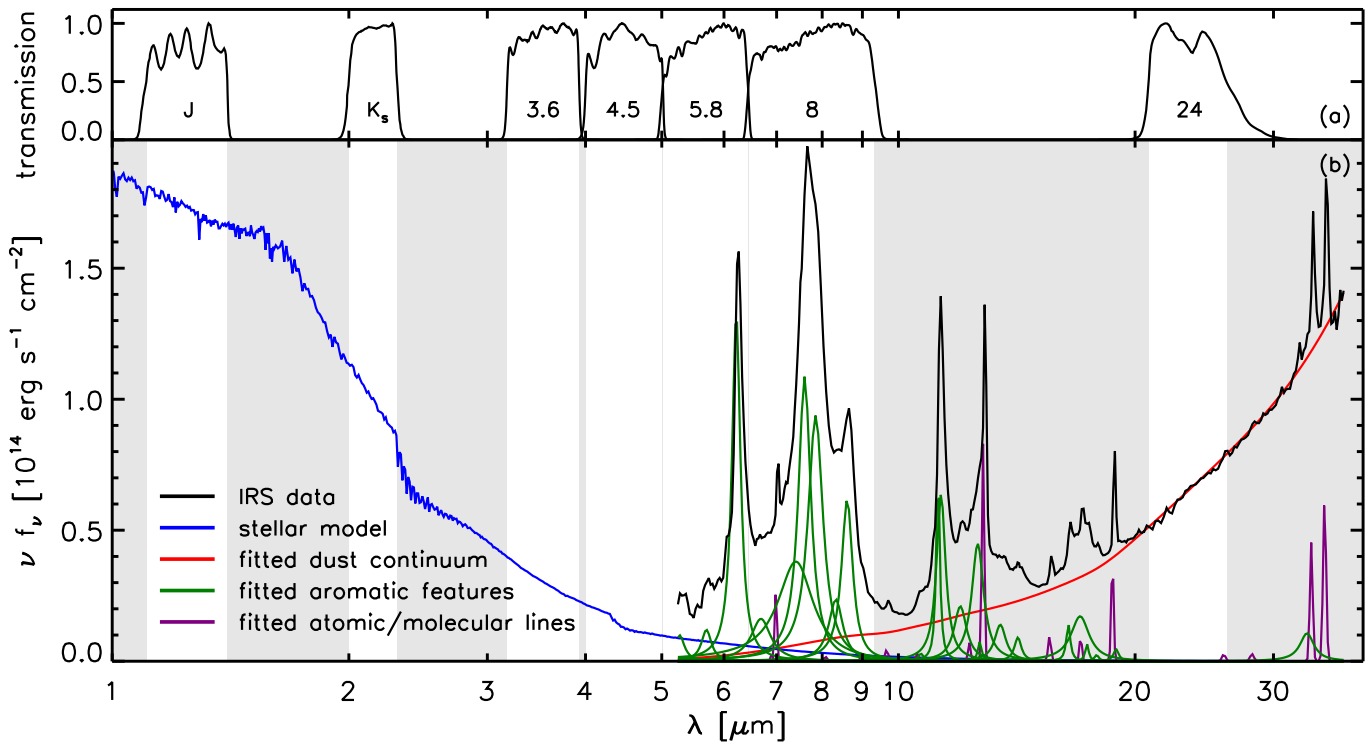}
\end{center}
\caption{
  (a) Spectral response of the $J$, $K_s$, IRAC 3.6/4.5/5.8/8.0 $\mum$,
  and MIPS $24\mum$ broad bands (the lighter regions below indicate wavelength ranges
  where the response
  is greater than half of the maximum). 
  (b) IRS spectrum of 
  NGC~4536 (solid black line) and its constituent flux contributions.  The stellar 
  model (blue) is an interpolation of the 100 Myr Starburst99 SEDs (for the metallicity of NGC~4536), extinction-corrected and 
  scaled to match the $J-K_s$ color and $3.6\mum$ flux of NGC~4536.  The 
  aromatic features (green), atomic/molecular lines (purple), and dust continuum (red) were 
  simultaneously fit using PAHFIT.}
  \label{fig_sed}
\end{figure*}

Investigations of mid-infrared aromatic feature emission are generally based on 
spectroscopic observations made from above the Earth's atmosphere; for example,
with \emph{Spitzer}'s Infrared Spectrograph \citep[IRS;][]{2004hou154apjs18}.
As such, they are largely relegated to samples of galaxies that are relatively 
small and/or bright.  Taking advantage of the fact that 
several of the strongest features fall within the rest-frame $8\mum$ band 
(see Figure~\ref{fig_sed}) of the Infrared
Array Camera \citep[IRAC;][]{2004faz154apjs10} on {\it Spitzer},
\citet{2005eng628apj29,2008eng678apj804} introduced a purely photometric approach
to measuring the strengths of aromatic features.  
However, the agreement between photometrically and spectroscopically
determined AFE equivalent widths for 27 starburst galaxies in 
\citet{2008eng678apj804} was characterized by a Spearman rank-order
correlation coefficient of only $r_s = 0.46$.

In this paper, we refine the photometric prescription for measuring 
aromatic feature emission at $8\mum$, 
demonstrate its improved reliability, and apply it to the 
statistically complete sample of 258 local star-forming galaxies comprising 
the Local Volume Legacy survey \citep[LVL;][]{2009dal703apj517}.  By doing so,
we hope to advance future AFE investigations 
with a larger, more diverse, and less biased data set.
In \S~\ref{sec_galaxies}, we introduce the LVL galaxies as well as an 
additional spectroscopic sample used for diagnostic purposes.  The 
corresponding datasets are described in \S~\ref{sec_data},
while in \S~\ref{sec_methods} our methodology is detailed and tested.  
Finally, we present an inventory of LVL aromatic 
feature emission measurements as well as a brief characterization of them
in \S~\ref{sec_lvlafe} before summarizing our findings in \S~\ref{sec_concl}.

\section{Galaxies}\label{sec_galaxies}

The \emph{Spitzer Space Telescope} Local Volume Legacy is a two-tiered 
survey of a statistically complete sample of nearby star-forming 
galaxies.  The inner tier includes all known galaxies within 3.5 Mpc that
lie outside the Local Group and the galactic plane ($|b| > 20\degr$), as well
as those galaxies in the M81 group and the Sculptor filament.
The outer tier is comprised of all galaxies brighter than $m_B < 15.5$ 
that lie within 11 Mpc and have $|b| > 30\degr$.  The resulting 258 galaxies 
are primarily those of the ACS Nearby Galaxy Survey Treasury
\citep[ANGST;][]{2009dal183apjs67} and the 11 Mpc H$\alpha$ and Ultraviolet
Galaxy Survey 
\citep[11HUGS;][]{2008ken178apjs247,2009lee706apj599}; however, a more detailed 
accounting can be found in \citet{2009dal703apj517}.  
The nearly volume-limited nature of LVL means that it is dominated in number by 
traditionally underrepresented low-luminosity dwarf galaxies, while the 
volume itself is sufficiently large to ensure a diverse 
cross-section of star formation properties and morphologies.

Overlapping with the 258 LVL galaxies are 40 SINGS 
\citep{2003ken115pasp928} galaxies with suitable mid-infrared spectroscopy for
testing and calibrating the photometric techniques applied in this paper.
The combined total of 283 nearby galaxies addressed in this study is listed 
in Table~\ref{tab_galaxies} by name and 
celestial coordinates, along with their redshifts, distances, oxygen abundances
compiled from the literature (see \S~\ref{sec_abun}), and 
inclusion in the LVL and/or spectroscopic subsamples.  
Additional properties 
of the LVL and SINGS galaxies which are not directly relevant to this 
body of work can be found in \citet{2009dal703apj517} and 
\citet{2003ken115pasp928}, respectively.

\section{Datasets}\label{sec_data}

\subsection{LVL Photometry}

Our analysis of aromatic emission in the LVL galaxies 
draws upon global
broadband photometry in the $J$ 
(1.25 $\mum$) and $K_s$ (2.17 $\mum$) bands, the four \emph{Spitzer} IRAC bands 
(3.6, 4.5, 5.8, and 8.0 $\mum$) and the three \emph{Spitzer} MIPS bands (24, 
70, and 160 $\mum$).  These data are presented in the prior
LVL survey description and infrared photometry paper \citep{2009dal703apj517},
where a detailed account is given of what is summarized here.

IRAC and MIPS observations were
taken in two epochs at different rotations in order to remove foreground 
asteroids and mitigate detector artifacts.  The total effective exposure time
is 240 seconds for the IRAC bands and 160, 80, and 16 seconds at 24, 70, and 
160 $\mum$, respectively.  For the brighter targets, additional 1.2 second IRAC 
images were used to recover pixels that were either saturated or in the 
non-linear regime.  Foreground stars and background galaxies were removed from
the images before photometry was extracted using the same elliptical aperture 
(chosen to include all detectable light) for each band.  The
final flux densities are corrected for Galactic extinction 
\citep{1998sch500apj525} assuming $A_V / E(B-V) \approx 3.1$ and the reddening
curve of \citet{2001li554apj778}, and include extended source aperture 
corrections but no color corrections (which are on the order of a few percent, but
are not relevant for this study; see \S~\ref{sec_afe}).  
The $J$ and $K_s$ band photometry was
similarly extracted from Two Micron All Sky Survey 
\citep[2MASS;][]{2006skr131aj1163} images using the same apertures and masking.
  
Nearly all of the LVL 
galaxies are detected in all seven \emph{Spitzer} bands down to $m_B\approx14$ 
mag and $\mathrm{M}_{B}\approx-13$ mag.  Where fainter galaxies were not detected, 
$5\sigma$ upper limits are provided.  This is the case for 59 galaxies in at
least one band and nine galaxies in all nine infrared bands.  Individual 
photometry is not provided for MCG~-05-13-004, but is instead included with
that of NGC~1800, as the two spatially overlap.  In addition, IC~5152 was not
observed at $3.6\mum$ or $5.8\mum$, $J$/$K_s$ photometry is not available 
for the LMC, and $5.8\mum$/$J$/$K_s$ data is not included for the SMC.

\subsection{LVL Oxygen Abundances}\label{sec_abun}

One of the expected key parameters for the formation and abundance of 
aromatic molecules is the metallicity of the interstellar medium.  
Thus, we have compiled oxygen abundances from the literature
(Table~\ref{tab_galaxies}), where measurements
were available for 129 of the 258 LVL galaxies.
Unless otherwise noted, all abundances are as given 
by the original source.  When multiple sources exist, the listed abundance is 
the one expected to be most representative of the metallicity throughout the 
star-forming disk. Specifically, for low mass galaxies, where oxygen abundance 
gradients are negligible 
\citep[e.g.,][]{1996kob471apj211,1997kob489apj636,2006van636apj214},
an average of high signal--to--noise ratio observations 
was adopted if multiple observations were available.
  For high mass galaxies, 
where abundance gradients can be substantial 
\citep[e.g.,][]{1994zar420apj87}, we adopt either the value from an 
integrated spectrum or the rough equivalent at $0.4\,\mathrm{R}_{25}$.

Of particular concern 
for oxygen abundance compilations is the wide range of empirical strong-line 
abundance calibrations available in the literature.  Systematic differences in 
empirical calibrations can result in abundance discrepancies of as much as 
0.7 dex 
\citep[see, e.g., the extensive discussion in][]{2008kew681apj1183}.
While the oxygen abundances tabulated in Table~\ref{tab_galaxies} 
have not been corrected 
for such systematic effects, the majority are on similar calibration scales 
with potential offsets of only 0.1 to 0.2 dex between sources 
\citep[e.g.,][]{1991mcg380apj140,1994zar420apj87}.
One further 
complication for empirical estimates of the oxygen abundance is the double 
valued nature of the strong-line abundance indicators (e.g., R23).  We 
revise the oxygen abundance for UGC~04787, NGC~3510, UGC~06900, NGC~4248, and 
UGC~07699 based on published line strengths and the assumption that they are 
likely low metallicity systems, rather than high metallicity as published 
in the original source.  
Finally, we tabulate the oxygen abundances published 
for KDG~61 and UGC~05336, but choose not to include them in our analysis 
(see \S~\ref{sec_afe_vs_oh})
as it is not clear that these measurements are indicative of their ISM
\citep{2009cro705apj723}.

\subsection{SINGS Spectroscopy}\label{sec_irs}

The galaxies comprising the spectroscopic sample were all spectrally mapped 
with \emph{Spitzer's} Infrared Spectrograph \citep[IRS;][]{2004hou154apjs18} as 
part of the SINGS survey.  Full observational details regarding this spectral 
mapping are provided in \citet{2003ken115pasp928}, while the final
one-dimensional spectra are presented in \citet{2007smi656apj770}.  In brief,
the latter span the wavelength range $5-38 \mum$ with spectral resolution
$R\approx60-125$ and correspond to extraction apertures (provided in Table~1 of
\citet{2007smi656apj770}) that were chosen to include the largest useful region of 
circumnuclear and inner disk emission (approximately the inner kiloparsec).  

The spectra are comprised of data from 
the four low-resolution orders of IRS (SL2: $5.25-7.6 \mum$, SL1: 
$7.5-14.5 \mum$, LL2: $14.5-20.75 \mum$, LL1: $20.5-38.5 \mum$) which were 
matched in the overlap regions in such a way that preserves the flux calibration
of the LL2 segment.  However, for the purposes of this study, they have been
normalized (with scale factors on the order of unity) to identically 
agree with the IRS aperture matched IRAC $8\mum$ flux density as described below in 
\S~\ref{sec_apm}.
The resulting spectra are shown in Figure~\ref{fig_spec1}, with their statistical 
uncertainties shaded.  However, as noted in \citet{2007smi656apj770}, 
at signal--to--noise greater than approximately 10, systematic 
errors exceed the statistical uncertainties in some regions.  Therefore, throughout
our analysis, we adopt uncertainties that floor the signal--to--noise at 10.

\subsection{Synthetic 2MASS/\emph{Spitzer} Photometry}

At various stages, our analysis employs synthetic photometry derived from
either observed or modeled spectra.  Here, ``synthetic'' refers to photometry
that \emph{would} have been obtained if a source with a spectral energy
distribution given by the spectrum in question had been imaged.  Essentially, this amounts to a
convolution of the spectrum with the appropriate response function for the 
desired photometric imager/band.  In the case of 2MASS $J$ and $K_s$, 
relative spectral response curves in the appropriate units to be
integrated directly over $\mathrm{f}_\lambda$ spectra are provided by \citet{2003coh126aj1090}.
A few more steps are involved in calculating synthetic photometry for the IRAC and 
MIPS bands; however, this has been automated in the IDL routine
{\tt SPITZER\_SYNTHPHOT} which is available
from the \emph{Spitzer} Science Center.  

\subsection{SINGS IRS Aperture Matched Photometry}\label{sec_apm}

Our purpose for the spectroscopic sample is to compare measurements derived 
from photometry with those obtained directly from spectroscopy.
In order to ensure meaningful comparisons, the IRAC and MIPS $24\mum$
SINGS photometry for the spectroscopic
sample was extracted with apertures matched to the IRS data.  Aside from this,
an improved background subtraction algorithm, 
and the aperture corrections described in \citet{2007smi119pasp1133},
it corresponds to the 
global photometry detailed in \citet{2007dal655apj863}, which closely
resembles the LVL photometry in its observational details and data reduction.

\begin{figure*}[t]
\includegraphics[width=6.81in]{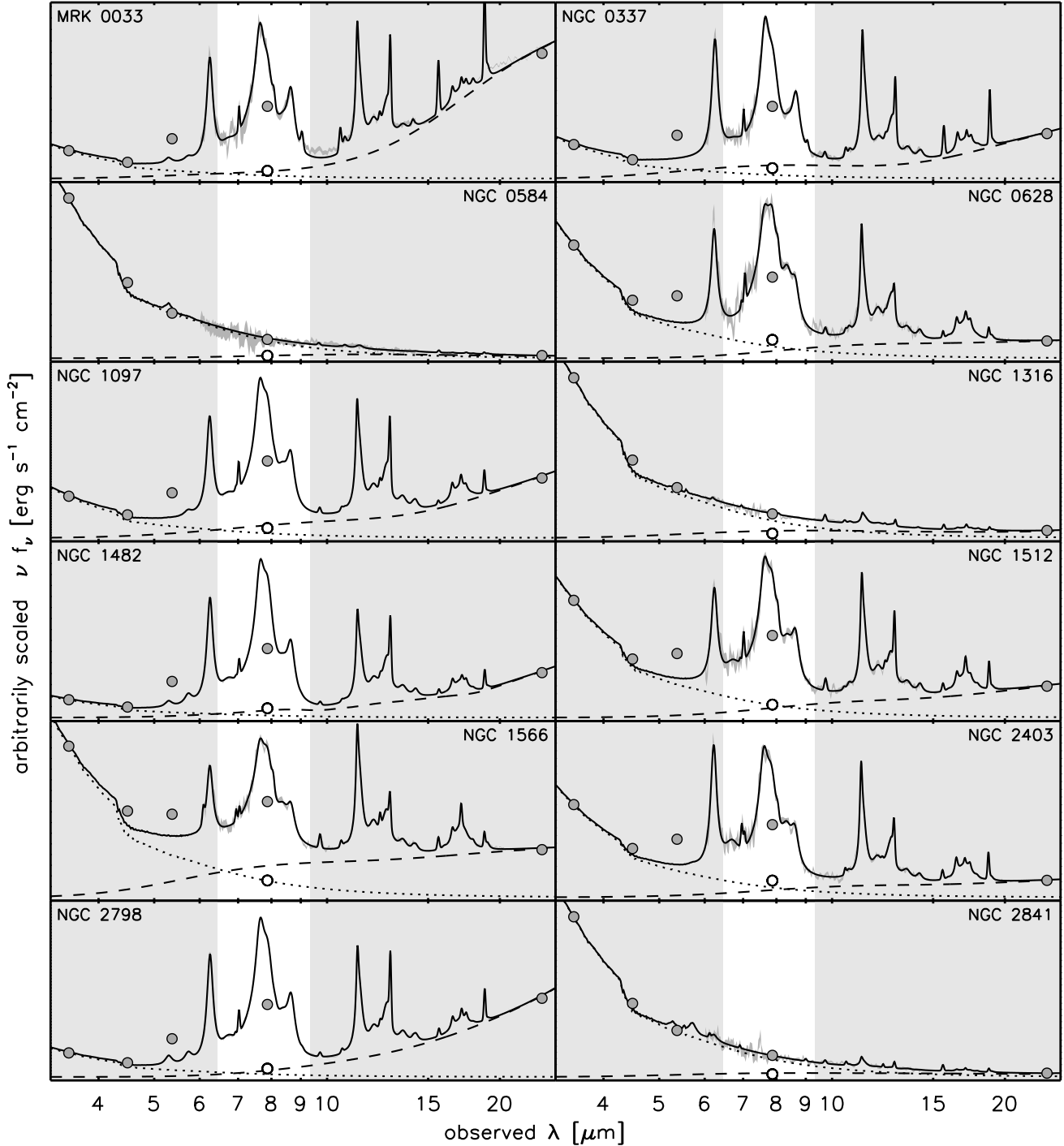}
\caption{\emph{Spitzer} IRS spectra (solid line) with aperture matched
  IRAC photometry and synthetic MIPS $24\mum$ photometry (filled circles)
  for the spectroscopic sample.  The unshaded portion of
  the logarithmic wavelength axis
  corresponds to the coverage of the 
  IRAC $8\mum$ band.  The 1-$\sigma$ envelope of the \emph{statistical} 
  uncertainty in the IRS spectra is also shaded.  The modelled stellar
  and fitted (using PAHFIT) dust continuum contributions are shown as dotted and dashed 
  lines, respectively.  The latter can be compared to the photometrically derived value 
  (empty circle) given by Equations~(\ref{eq_powerlaw})$-$(\ref{eq_c3}).}
  \label{fig_spec1}
\end{figure*}

Any residual calibration differences between the SINGS spectroscopy and 
photometry were removed by scaling the IRS 
spectra such that synthetic $8\mum$ photometry identically reproduced 
the aperture matched IRAC $8\mum$ flux density.  This introduces greater 
uncertainty than normalization to the MIPS $24\mum$ flux density, and 
would be inadvisable if absolute measurements of AFE were our primary
concern.  Since we are instead motivated by minimizing relative 
differences between the spectroscopic and photometric values, it is 
to our advantage to scale the IRS spectra to match the site of the 
aromatic feature complex of interest.
Furthermore, synthetic $24\mum$ photometry was obtained directly from the 
scaled spectra, in lieu of the aperture matched MIPS values.
The rms of the corresponding relative difference is approximately 10\%,  
which we universally adopt as the 
uncertainty in the broadband photometry used for the spectroscopic
sample (provided in Table~\ref{tab_spectroscopic}).

\begin{figure*}[t]
\setcounter{figure}{1}
\includegraphics[width=6.81in]{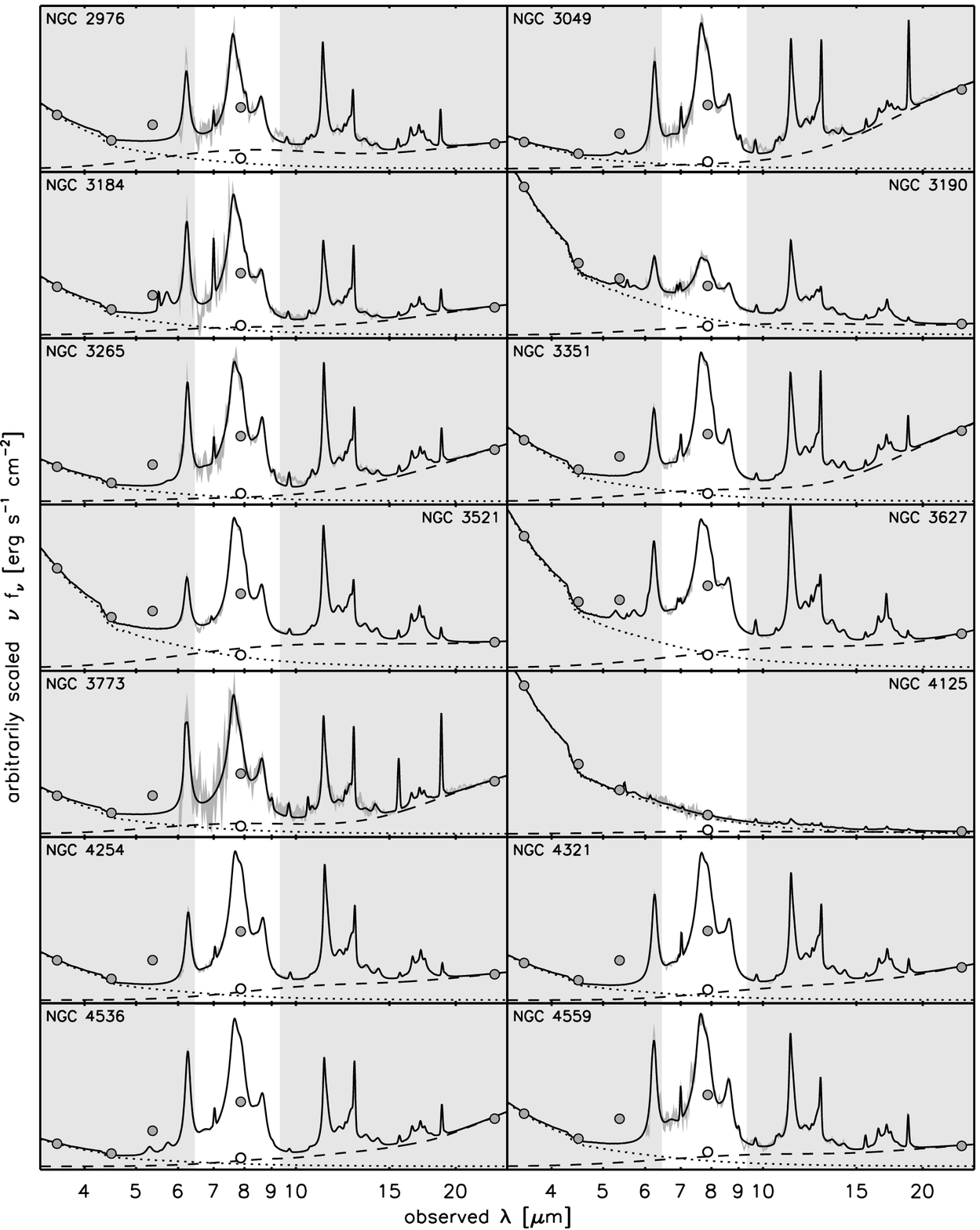}\\[5mm]
\centerline{FIG. \ref{fig_spec1}.---\emph{Continued}}
\end{figure*}

\begin{figure*}[t]
\setcounter{figure}{1}
\includegraphics[width=6.81in]{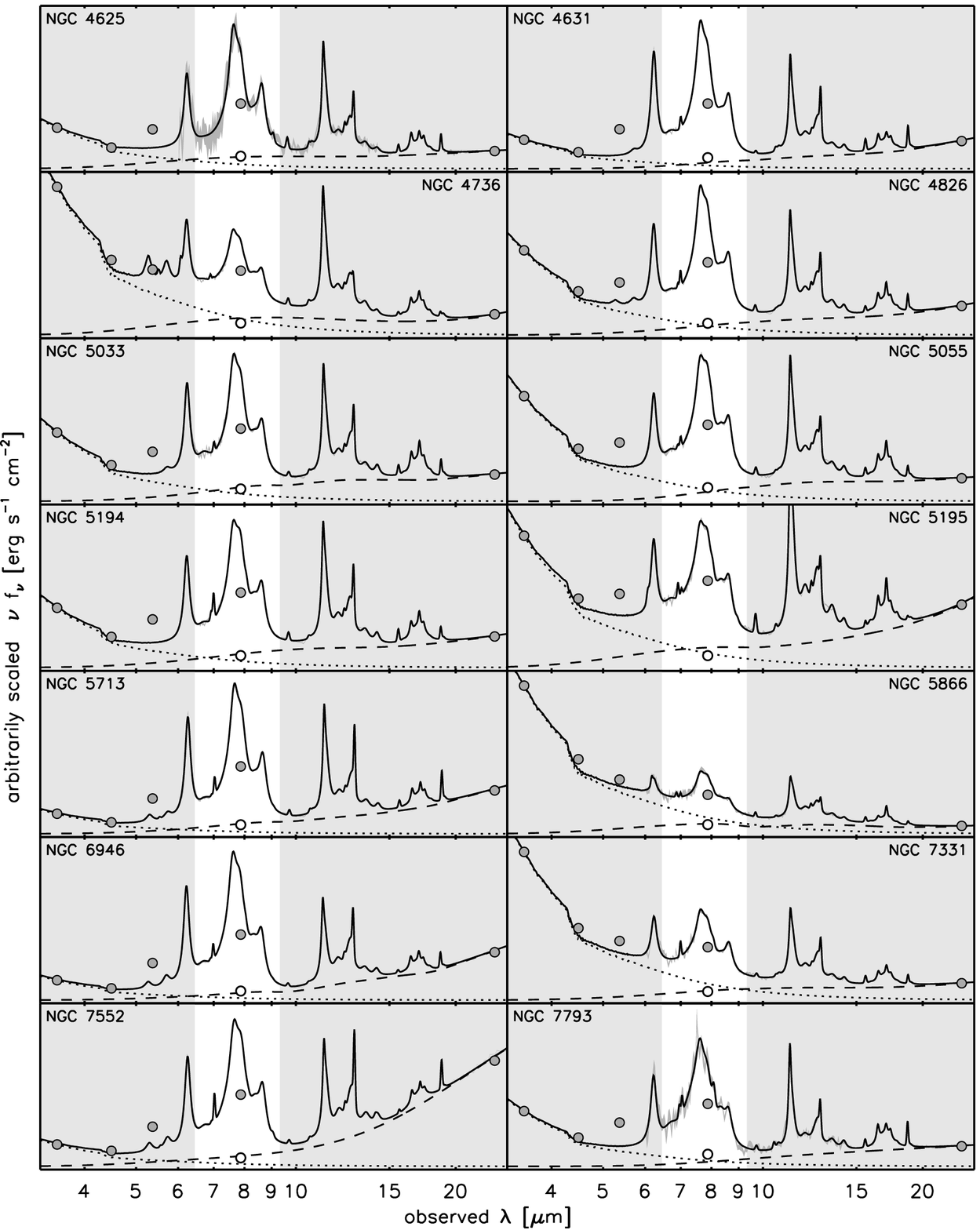}\\[5mm]
\centerline{FIG. \ref{fig_spec1}.---\emph{Continued}}
\end{figure*}

\clearpage 

\section{Methodology}\label{sec_methods}

\subsection{LVL L-Z Relationship}\label{sec_lz}

The LVL oxygen abundances compiled from the literature (see \S~\ref{sec_abun})
exhibit the
well known luminosity-metallicity (L-Z) correlation with stellar luminosity 
(Figure~\ref{fig_lvl_oh_vs_i2}).
Weighted by the statistical 1-$\sigma$ uncertainties for both parameters,
the best linear fit is given by
\begin{equation}\label{eq_oh_vs_i2}
  12+\log\mathrm{(O/H)} = 5.06\pm0.04 - (0.164\pm0.002)\,\mathrm{M}_{4.5},
\end{equation}
which agrees reasonably well with previous results for other samples
\citep[e.g.,][]{2006lee647apj970}.  The scatter in oxygen abundance with 
respect to the linear fit is
generally consistent with the statistical uncertainties plus an additional 
approximately 0.15 dex systematic calibration error.
A few outliers (e.g., UGC~05340 and NGC~4656) may also be affected by
errors in distance measurements.  Subsequently, we adopt the oxygen abundance
given by Equation~(\ref{eq_oh_vs_i2}) for galaxies without measurements in the
literature.  However, these L-Z based abundances are used only for the purpose of
applying metallicity corrections to stellar SED models (\S~\ref{sec_star}).

\subsection{AFE Measurement}\label{sec_afe}

The rest-frame mid-infrared portion of a galaxy's spectral energy 
distribution is typically a combination of aromatic features, atomic and 
molecular lines, and continua of dust emission and starlight.  As 
demonstrated by \citet{2007smi656apj770} and illustrated in
Figure~\ref{fig_sed}, these components can be simultaneously 
fitted given a spectrum of sufficient resolution and signal--to--noise.  
One measure of the strength of the aromatic emission, $\FIrsIivAfe$, is then 
given by integrating and summing the flux in the $8\mum$ complex formed 
by individual features at 7.42, 7.60, 7.85, 8.33, and 8.61 $\mum$.

\begin{figure}[b]
\setcounter{figure}{2}
\begin{center}
\includegraphics[width=3.3in]{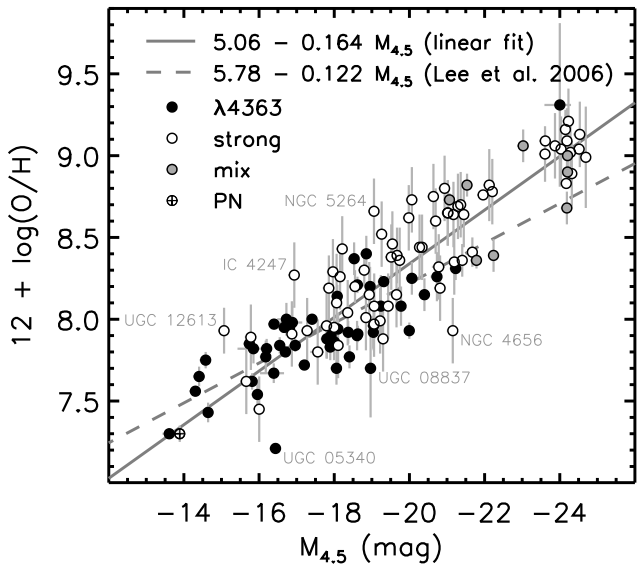}
\end{center}
\caption{Oxygen abundance versus IRAC $4.5\mum$ absolute magnitude for 
  127 LVL galaxies.
  Weighted by the statistical 1-$\sigma$ uncertainties shown for both parameters,
  the corresponding luminosity-metallicity relationship has a 
  $5.06\pm0.04 - (0.164\pm0.002)\,\mathrm{M}_{4.5}$ linear fit (solid line) 
  that is marginally consistent with the $5.78\pm0.21 - (0.122\pm0.012)\,\mathrm{M}_{4.5}$
  dashed line reported by \citet{2006lee647apj970}.  
  Note that some of the
  scatter (a few outliers are labeled) 
  is due to a \emph{systematic} uncertainty of $0.1-0.2$ 
  for the inhomogeneous oxygen abundance measurements drawn from the literature.}
\label{fig_lvl_oh_vs_i2}
\end{figure}

However, this complex conveniently falls within the broad IRAC $8\mum$ band,
providing an alternative means for measuring the strength of the aromatic 
features if their contribution can be disentangled from the total flux.  
For the spectroscopic sample, $3-96$ (with a mean of 19) percent of this 
total comes from starlight, while the dust continuum is responsible for 
$6-32$ (with a mean of 16) percent of the non-stellar emission.  Combined,
the two continuum sources contribute $12-100$ (with a mean of 34) percent of the IRAC
$8\mum$ flux density.  We build upon earlier work by
\citet{2005eng628apj29,2008eng678apj804}, exploiting our spectroscopic galaxy
sample to motivate, calibrate, and test a purely photometric determination
of aromatic emission at $8\mum$, $\FIivAfe$.
For clarity, spectroscopically derived values are distinguished from their
photometrically based counterparts by a parenthetical letter S in the subscript
(e.g., $\FIrsIivAfe$).

\subsubsection{Starlight}\label{sec_star}

The same determination of the stellar contribution to the mid-infrared
is utilized for both the spectroscopic and photometric analyses; 
therefore, we address it separately and first.
The stellar continuum is the tail of a distribution that peaks in the 
optical or near-infrared, which, like \citet{2008eng678apj804}, we model
using SEDs generated by the stellar population synthesis code 
Starburst99 and described in \citet{1999lei123apjs3} and 
\citet{2005vaz621apj695}.
These are available for a range of ages between one Myr and one Gyr and
for metallicities $\mathrm{Z}=0.040, 0.020, 0.008, 0.004, \mathrm{and}\ 0.001$,
or $12+\log\mathrm{(O/H)} = 7.6, 8.2, 8.5, 8.9, \mathrm{and}\ 9.2$ 
given the Starburst99 assumed solar values of $\mathrm{Z}_\sun=0.02$ and 
$12+\log\mathrm{(O/H)}_\sun=8.9$.

As confirmed by \citet{2008eng678apj804}, changes in the stellar
continuum slope for these instantaneous star formation models are
insignificant in the IRAC bands beyond $5-15$ Myr; therefore, we
arbitrarily adopt an age of 100 Myr.  The corresponding set of SEDs
are redshifted and interpolated to match either the observed velocity
and oxygen abundance listed in Table~\ref{tab_galaxies} or, if
unavailable, the values approximated by the Hubble relation
($\mathrm{v_{km/s}} = 72\,\mathrm{D_{Mpc}}$) and
Equation~(\ref{eq_oh_vs_i2}).  The resulting spectrum is adjusted for
internal extinction, assuming the extinction law of \citet{1985rie288apj618}
and a simple foreground screen, by multiplying it with
$10^{-0.1516\,A_V\,\lambda^{-1.475}}$, where $A_V$ is equal to $16.04
\left(\log\left(\mathrm{f}^{\,\mathrm{sb99}}_{\,\mathrm{J}}/
                \mathrm{f}^{\,\mathrm{sb99}}_{\,\mathrm{K_s}}\right) -
\log\left(\mathrm{f}^{\,\mathrm{obs}}_{\,\mathrm{J}} /
          \mathrm{f}^{\,\mathrm{obs}}_{\,\mathrm{K_s}}\right)\right)$, 
$\mathrm{f}^{\,\mathrm{sb99}}$ corresponds to synthetic
photometry from the stellar spectrum, and $\mathrm{f}^{\mathrm{obs}}$ is the observed
photometry for the galaxy in question. In the case of 39 LVL galaxies,
$\mathrm{f}^{\,\mathrm{sb99}}_{\,\mathrm{J}}/
 \mathrm{f}^{\,\mathrm{sb99}}_{\,\mathrm{K_s}}$ is unknown and 1.22 (the median of
the remainder) is assumed.  For the galaxies considered in this paper,
these redshift, metallicity, and extinction corrections are a few percent
or less.

Whereas \citet{2008eng678apj804} normalized their stellar SED to the mean of
the observed $K_s$ and IRAC $3.6\mum$ band flux densities, we choose to use
only the latter.  This has the advantage
that the absolute calibration of our methodology is 
tied to a single photometric system, not to mention the fact that
available $K_s$ band photometry is highly uncertain for
the dwarf galaxies comprising the majority of the LVL sample.
This normalization is done by
scaling the stellar model such that synthetic IRAC photometry matches
the observed photometry at $3.6\mum$.  Implicit here is the assumption that
the IRAC $3.6\mum$ band is a reliable tracer of global stellar emission, an assertion
supported by \citet{2004pah154apjs235} and the modelling of \citet{2007dra657apj810}.
As we are not addressing starburst galaxies or those with especially
hot dust, we do not need to be concerned with contamination from 
a strong dust continuum or 
significant aromatic emission from the $3.3\mum$ feature.

Determining the stellar component of the IRS spectra described in \S~\ref{sec_irs} is 
then simply a matter of interpolating the stellar SED onto the same wavelength
scale.  Similarly, synthetic photometry of the stellar SED yields
the stellar component of the flux density in band X (henceforth referred to as $\fXStr$).
The resulting ratio of $\fXStr / \fIiStr$
is approximately 58, 39, 24, and 2.8 percent, respectively, for X = 4.5, 5.7,
8, and 24 microns.  
Table~\ref{tab_spectroscopic} provides $\fIivStr$ for the spectroscopic sample,  
while Table~\ref{tab_photometric} includes $\fXStr$ for the MIPS $24\mum$ band and all of the IRAC bands
(except for $3.6\mum$ where $\fIiStr = \fIi$).
Throughout the remainder of the paper, we represent the 
non-stellar, or stellar-subtracted, photometry with the $^{\mathrm{ns}}$ superscript,

\begin{equation}\label{eq_fNs}
\fXNs = \fX - \fXStr.
\end{equation}

\noindent Disregarding any error in the model or response function, the statistical 
uncertainty in our synthetic stellar photometry is given by 
appropriately scaling the uncertainty at $3.6\mum$ flux.  
For the data considered in this 
paper, the resulting uncertainty in the stellar flux is always less than the
corresponding uncertainty in the total flux.  Thus, the error associated with
stellar subtraction is negligible.

\subsubsection{Spectroscopic Dust Decomposition}\label{sec_afespec}

Spectroscopic aromatic feature emission measurements were obtained from 
the stellar-subtracted IRS spectra 
using the IDL program {\tt PAHFIT} \citep{2007smi656apj770}.
This spectral decomposition code simultaneously fits up to eight dust continuum 
components with modified blackbodies at fixed temperatures ranging from
$35-300$ K,
18 Gaussian atomic and molecular lines (listed in Table~2 of 
\citet{2007smi656apj770}), 25 separate aromatic features with Drude 
profiles (listed in Table~3 of \citet{2007smi656apj770}), and absorption
from dust extinction in the form of a power law plus silicate features 
at 9.7 and 18 $\mum$.  
By default, {\tt PAHFIT} also fits a simple
starlight component
(a $5\times10^{3}\,\mathrm{K}$ blackbody); however, 
we fixed this to be zero for our previously stellar-subtracted spectra.

In addition to the spectroscopy, 
the stellar-subtracted IRAC 3.6 and 4.5\ $\mum$ photometry were included in the fit 
to extend the wavelength coverage at the blue end.
The results are shown in Figure~\ref{fig_spec1} and
quantified in Table~\ref{tab_spectroscopic} in the form
of synthetic IRAC $8\mum$ photometry of the 
atomic/molecular lines 
($\fIrsIivAmo$), total dust continuum ($\fIrsIivCnt$), and 
$8\mum$ complex of aromatic features (7.42, 7.60, 7.85, 8.33, and 8.62 $\mum$;
$\fIrsIivAfe$), as well as integrated fluxes for the $8\mum$ complex ($\FIrsIivAfe$)
and the 23 aromatic features spanning the mid-infrared from $5.5\mum$ to $20\mum$
($\FIrsMirAfe$). Synthetic IRAC $8\mum$ photometry of the
previously subtracted stellar continuum ($\fIivStr$) is also provided.

\subsubsection{Photometric Dust Decomposition}\label{sec_afephot}

The photometric equivalent of the integrated flux in the $8\mum$ complex
of aromatic features is proportional to the contribution of those 
features to the monochromatic IRAC $8\mum$ flux density $\fIivAfe$, 

\begin{equation}\label{eq_FIivAfe}
\FIivAfe = c_1 \ 10^{-23} \ \fIivAfe.
\end{equation}

\noindent The constant $c_1$ (to be solved for subsequently) primarily reflects the width of the IRAC $8\mum$ 
transmission curve ($\Delta\nu\approx1.3\times10^{13}$ Hz), but also, 
to first order, its departure from an idealized step function.  The factor 
of $10^{-23}$ is simply for convenience, yielding units of 
$\mathrm{erg}\,\mathrm{s}^{-1}\,\mathrm{cm}^{-2}$ for $\FIivAfe$ given $\fIivAfe$
in Janskys.

We define $\fIivAfe$ to be the scaled remainder of the non-stellar IRAC $8\mum$ flux density
after subtracting the contribution from the dust continuum,

\begin{equation}\label{eq_fIivAfe}
\fIivAfe = c_2 \left(\fIivNs - \fIivCnt \right).
\end{equation}

\noindent The scale factor ($c_2 \lesssim 1$) generically accounts for 
additional aromatic features that overlap with the IRAC $8\mum$ band and, to a 
much lesser degree, flux from atomic and molecular lines.  Both of these
contributions are indeed correlated to $\fIivAfe$.

We estimate the dust continuum contribution to the IRAC $8\mum$ band by 
extrapolating a power law tethered to stellar-subtracted neighboring bands. 
At the red end, MIPS $24\mum$ is conveniently dominated by the dust 
continuum, despite being further removed from $8\mum$ than would be ideal.
Several choices for the blue end are provided by the remaining IRAC bands.
The $5.8\mum$ band has the disadvantage of including a strong aromatic feature 
($\lambda=6.22\mum$), whereas the $4.5\mum$ band can have insufficient
dust continuum flux for reliable anchorage of the power law.  As a compromise,
we adopt the mean of the two,

\begin{equation}\label{eq_powerlaw}
\fIivCnt = c_3 \left( \frac{\fIiiNs + \fIiiiNs}{2} \right)^{1-\alpha} {\fMiNs}^{\,\alpha},
\end{equation}

\noindent where

\begin{equation}\label{eq_alpha}
\alpha = \frac{\log \left(\lambda_{\,8}\right)  - \log \left( \frac{\lambda_{\,\mathrm{4.5}} + \lambda_{\,\mathrm{5.8}}}{2} \right)}
              {\log \left(\lambda_{\,24}\right) - \log \left( \frac{\lambda_{\,\mathrm{4.5}} + \lambda_{\,\mathrm{5.8}}}{2} \right)} = 0.282,
\end{equation}

\noindent and $\lambda_{\,\mathrm{X}}$ refers to the effective wavelength of band X
($\lambda_{\,\mathrm{4.5}} = 4.493\mum$, $\lambda_{\,\mathrm{5.8}} = 5.731\mum$, $\lambda_{\,\mathrm{8}} = 7.872\mum$, and  
$\lambda_{\,\mathrm{24}} = 23.675\mum$).

\begin{figure}[t] 
\begin{center}
\includegraphics[width=3.3in]{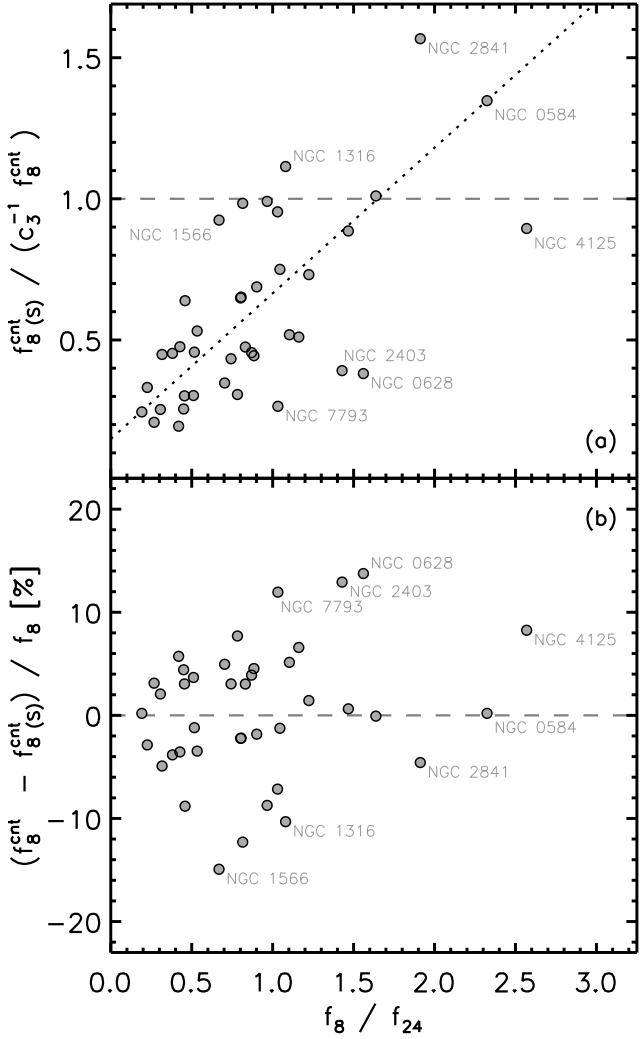}
\end{center}
\caption{(a) The ratio between the spectroscopically determined dust
  continuum contribution to the IRAC $8\mum$ band ($\fIrsIivCnt$) and
  the photometric value given by Equations~(\ref{eq_powerlaw}) and 
  (\ref{eq_alpha}) with $c_3=1$ is correlated to the IRAC $8\mum$ over
  MIPS $24\mum$ color in the spectroscopic sample of SINGS galaxies
  and systematically differs from unity (dashed line). 
  (b) Adopting the linear fit from (a), $c_3 = 0.149 + 0.516\,\mathrm{f}_8 / \mathrm{f}_{24}$ 
  (dotted line),
  the difference between the photometrically determined dust continuum
  flux ($\fIivCnt$) and the spectroscopic value (relative to the total
  IRAC $8\mum$ flux) is scattered about zero (dashed line) with an rms of 0.064.}
  \label{fig_dust}
\end{figure}

The leading coefficient $c_3$ is included to account for any
systematic difference between our simple model and the true dust continua.
A constant value of 0.594 yields the same mean for 
$\fIivCnt$
as the synthetic $8\mum$ photometry from the fitted dust continua;
however, the success of
the power law indicator in reproducing $\fIrsIivCnt$ for a given galaxy
is correlated with mid-infrared color 
(Figure~\ref{fig_dust}a), and the rms of the difference relative to
the total flux is improved by adopting

\begin{equation}\label{eq_c3}
c_3 = 0.149 + 0.516 \left( \frac{\fIiv}{\fMi} \right).
\end{equation}

With this correction applied, the rms of the relative error in $\fIivCnt$
is still 0.54.  However, for the purposes of measuring aromatic emission,
we are primarily concerned with how well we can constrain the fraction of
$\fIiv$ contributed by the dust continuum.  Together, 
Equations~(\ref{eq_powerlaw})$-$(\ref{eq_c3}) recover
the spectroscopically determined fraction $\fIrsIivCnt/\fIiv$ with an rms 
of only 0.064 (see Figure~\ref{fig_dust}b, as well as individual results
included in Figure~\ref{fig_spec1}).
While power law extrapolation
is not the only means for estimating the dust continuum, other methods
(e.g., spectral templates) bear little hope for producing
better overall agreement if they are calibrated 
with the same free parameters (photometry). 
Future datasets including flux densities
at more ideal wavelengths may improve these results or enable a more
sophisticated methodology.

Combining Equations (\ref{eq_fIivAfe}) through (\ref{eq_c3}) allows for the determination
of $c_2$.  Minimizing the rms of the error 
(i.e., the difference between the photometrically and 
spectroscopically determined values) in the fraction 
$\fIivAfe / \fIiv$ yields $c_2 = 0.934$.
Likewise, adopting $c_1 = 1.53\times10^{13}$ in Equation~(\ref{eq_FIivAfe})
achieves the lowest rms for 
$(\FIivAfe - \FIrsIivAfe)\,/\,\FIiv$,
where $\FIiv \approx c_1 \ 10^{-23} \ \fIiv$.
Thus, our photometric prescription for the integrated flux in the five aromatic
features we refer to as the $8\mum$ complex, calibrated with the spectroscopic
sample, becomes

\begin{eqnarray}
  \FIivAfe =
    1.43 \times 10^{-10}\ \Big(\fIivNs & - & 
      ( 0.091 + 0.314\ \fIiv / \fMi ) \nonumber \\
    & \times & ( \fIiiNs + \fIiiiNs ) ^ {0.718} 
    \, {\fMiNs}^{\,0.282} \Big), \label{eq_FIivAfe_final}
\end{eqnarray}

\noindent where all of the flux densities are in Janskys and $\FIivAfe$ 
has units of $\mathrm{erg}\,\mathrm{s}^{-1}\,\mathrm{cm}^{-2}$.

\subsubsection{Photometric vs. Spectroscopic AFE}\label{sec_specvsphot}

\begin{figure}[t]
\begin{center}
\includegraphics[width=3.3in]{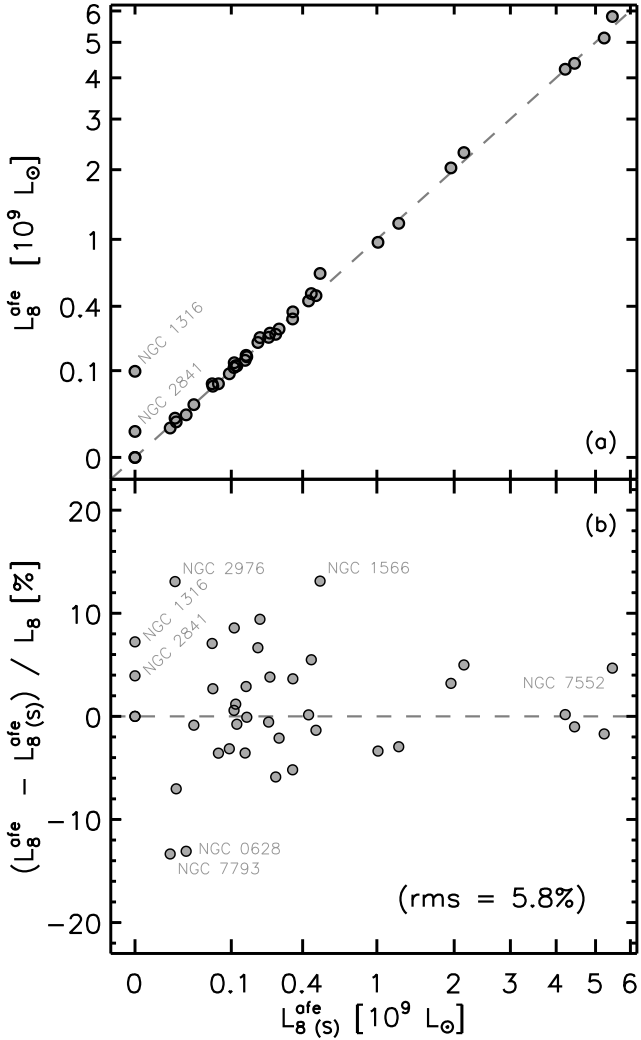}
\end{center}
\caption{(a) For the 40 SINGS galaxies in the spectroscopic sample,
  the photometrically determined AFE luminosities given by 
  Equation~(\ref{eq_FIivAfe_final}) recover the spectroscopic
  values well (the dashed line indicates equality), 
  with the two apparent outliers corresponding to galaxies
  with approximately zero aromatic emission.  As shown in (b), the
  maximal difference relative to the total $8\mum$ luminosity is
  12\% with a total rms about zero (dashed line) of 6\%.}  
  \label{fig_afe}
\end{figure}

Figure~\ref{fig_afe} demonstrates the agreement between the photometrically
determined $\FIivAfe$ from Equation~(\ref{eq_FIivAfe_final}) and the 
spectroscopically derived $\FIrsIivAfe$ from Table~\ref{tab_spectroscopic} for 
the SINGS galaxies comprising the spectroscopic sample.  Here the 
integrated fluxes have been converted to luminosities using the distances
provided in Table~\ref{tab_galaxies}.  The lower panel shows that the difference
in the methodologies relative to the total luminosity in the $8\mum$ band is 
at worst a 12\% effect (including the two apparent outliers in the upper panel 
which correspond to galaxies with approximately zero aromatic emission)
that improves with increasing luminosity in the aromatic
features.  The rms about zero (for the full luminosity range) is 6\%. 
This systematic error is dominated by uncertainty in the dust continuum at $8\mum$.  

Individual error bars for the quantity $(\LIivAfe - \LIrsIivAfe)\,/\,\LIiv$
were determined both by mathematically propagating the various photometric 
uncertainties through Equation~(\ref{eq_FIivAfe_final}) and by 
repeating the calculations 1000 times with Monte Carlo resampling of the 
uncertainties.  Both methods were in agreement, and both significantly 
exceeded the rms quoted above.  
This is, perhaps, not surprising as 
scaling the spectra to match the aperture matched photometry eliminated some degree
of systematic error from sources such
as tilts and/or offsets in the spectra and uncertain
extended source corrections in the photometry.  
Thus, 
we adopt the $0.06\,\FIiv$ systematic error as a representative uncertainty
in the measurement of $\FIivAfe$.  
Note, however, that this pertains to the methodology itself
and due care should be taken when applying it to data with significantly 
greater statistical uncertainties.  

Similarly, consideration of the galaxies on which these results 
are based should be taken into account when applying
this methodology to other samples.  The estimation of the dust continuum 
at $8\mum$ is a model-independent
extrapolation of a galaxy's SED and has been shown to be reliable
for both dust dominated SEDs with strong aromatic features and stellar 
dominated SEDs with little to no aromatic emission.  Thus, these results
are somewhat insensitive to a galaxy's particular properties.  
However, as an example, in cases of galaxies with very hot dust 
\citep[e.g., the starburst galaxy SBS~0335$-$052E;][]{2008eng678apj804},
the color correction given 
in Equation~(\ref{eq_c3}) may be insufficient, resulting in an underestimated dust
continuum and a corresponding error in the strength of the aromatic features
surpassing that found in this sample.

\subsubsection{Alternative Non-Stellar Scaling}

\begin{figure}[t]
\begin{center}
\includegraphics[width=3.3in]{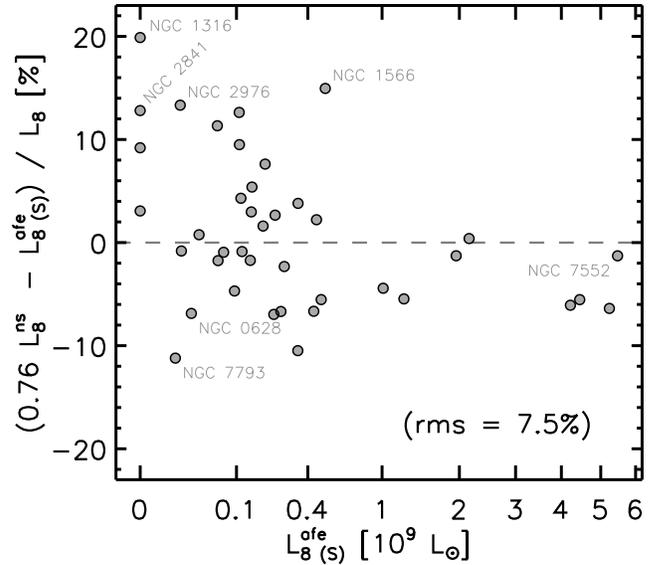}
\end{center}
\caption{Same as Figure~\ref{fig_afe}b, except with photometrically determined
  AFE luminosities given by 0.76 
  ($= c_2\,(1-0.19)$) times the non-stellar IRAC $8\mum$ luminosity
  instead of Equation~(\ref{eq_FIivAfe_final}).  The resulting agreement with
  the spectroscopic values is generally poorer, although comparable.  There is 
  a slight apparent luminosity trend, and, not surprisingly, the simple scaling 
  of the non-stellar light introduces greater errors for those galaxies with 
  very little relative aromatic emission (e.g., NGC~1316).}
\label{fig_nsafe}
\end{figure}

The dust decomposition of the stellar-subtracted IRAC $8\mum$ flux density given 
by Equation~(\ref{eq_powerlaw}) and discussed in \S~\ref{sec_afephot} is a 
simple power-law extrapolation of the dust continuum at 4.5 and 24 microns.  It 
is worth noting that an even simpler alternative is a multiplicative scaling of
the non-stellar light.  For the spectroscopic sample, Figure~\ref{fig_nsafe} 
shows results that are generally
comparable to Figure~\ref{fig_afe}b when $\fIivCnt$ is instead taken to be 
$\fIivNs \times 0.19$ (the mean from the PAHFIT values in 
Table~\ref{tab_spectroscopic}).  Not surprisingly, the disagreement between the 
photometric and spectroscopic AFE measurements is marginally poorer in most cases 
(the overall rms increases from 5.8\% to 7.5\%) and more signigicantly so for
galaxies with relatively weak or absent aromatic features (e.g., NGC~2841 or 
NGC~1316).  Scaling the non-stellar $8\mum$ emission has the advantage of not 
requiring additional adjacent photometry; however, we caution that systematic 
errors for 
galaxies with unusual SEDs \citep[e.g.,][]{2004hou154apjs211,
2007wu662apj952} will be increased as a result.

\subsubsection{Future Improvement}

\begin{figure}[b]
\begin{center}
\includegraphics[width=3.3in]{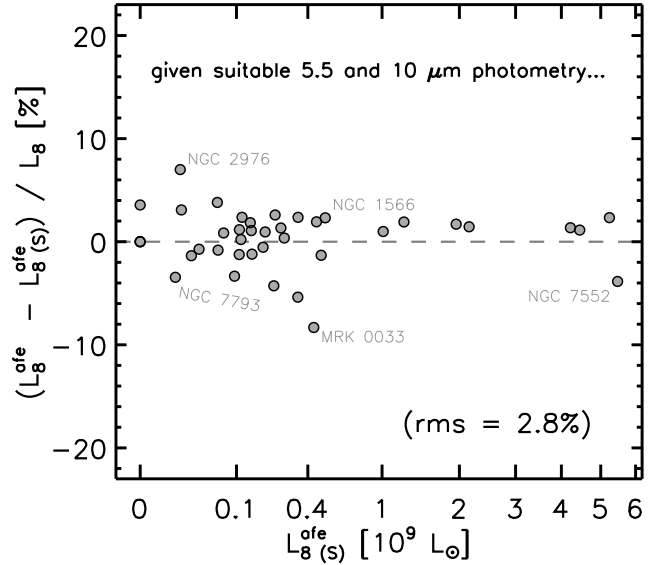}
\end{center}
\caption{Same as Figure~\ref{fig_afe}b, except with dust continuum measurements
  based on synthetic photometry at $5.5\mum$ and $10\mum$, demonstrating 
  that future datasets (e.g., from JWST) could further improve the 
  reliability of photometric mid-infrared aromatic feature emission
  measurements by a factor of two.}
  \label{fig_afe_future}
\end{figure}

Given the coarse approach of approximating a complex of spectral features
with broad-band photometry, the systematic uncertainty quoted in 
\S~\ref{sec_specvsphot} is impressive.  The most efficient means of improving
upon these results is more accurately constraining the dust continuum.  For 
example, Figure~\ref{fig_afe_future} demonstrates a 50\% reduction in scatter 
between the photometric
and spectroscopic AFE measurements, given a $5.5\mum$ band that avoids the 
aromatic feature at $\lambda=6.2\mum$ and a $10\mum$ band that samples
the continuum closer to the $8\mum$ complex than the MIPS $24\mum$ band.  
These, or complementary, data will be available in the future from 
the yet to be launched \emph{James Webb Space Telescope} \citep{2006gar123ssr485}.

\subsubsection{Mitigated Uncertainties}\label{sec_sum}

In \S~\ref{sec_specvsphot} we quote a modest systematic error in the photometric
measurement of aromatic emission in the $8\mum$ complex.
However, by construction, this uncertainty is essentially
random and thus mitigated when considering sufficiently large samples of 
galaxies.  For example, summing the spectroscopically and photometrically 
determined AFE values for the 40 SINGS galaxies in the spectroscopic sample 
respectively yields 
$\sum \LIrsIivAfe = 3.10\times10^{10}\,\Lsun$ and 
$\sum \LIivAfe = 3.14\times10^{10}\,\Lsun$, 
which differ by only 1.3\%.\\[5mm]

\begin{figure}[t]
\begin{center}
\includegraphics[width=3.3in]{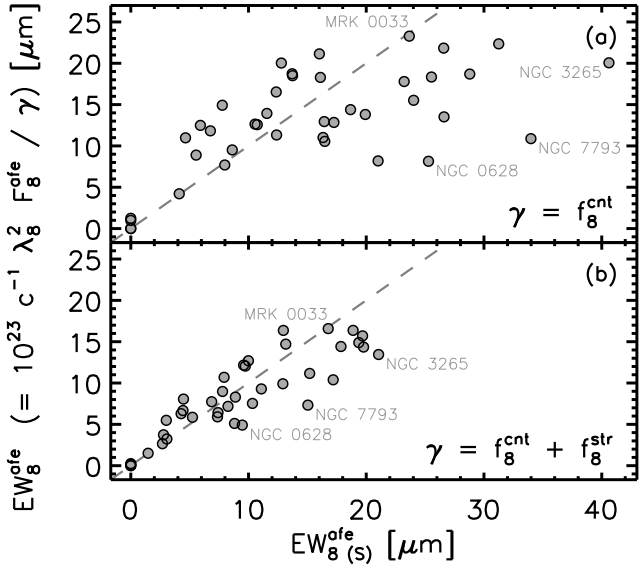}
\end{center}
\caption{Parametrizing the photometrically determined AFE values 
  as equivalent widths is inadvisable, as the agreement between the photometric
  and spectroscopic values seen in 
  Figure~\ref{fig_afe} is significantly diminished when dividing by
  either the relatively uncertain dust continuum (a) or the stellar plus dust continuum (b).}
  \label{fig_ew}
\end{figure}

\subsubsection{Equivalent Widths}\label{sec_ew}

In order to make comparisons between 
galaxies that are free of distance and luminosity biases,
line fluxes are often converted to equivalent widths (EW) by dividing them 
by the underlying continuum.  However, we 
caution that the agreement between the spectroscopic and photometric
methodologies decreases 
significantly if equivalent widths are used for the latter 
(compare Figure~\ref{fig_afe}a to 
Figure~\ref{fig_ew}).  This is precisely because the dust
continuum is the dominant source of uncertainty in $\FIivAfe$ and can itself
be relatively weak.  The scatter between 
the photometric $\mathrm{EW}^{\,\mathrm{afe}}_{\,8}$ and the spectroscopic 
$\mathrm{EW}^{\,\mathrm{afe}}_{\,8\,(S)}$
arises in part because an underestimated continuum results in an overestimated
aromatic emission strength, doubly inflating the equivalent width, and vice versa.
If a distance and luminosity independent measure of the aromatic feature emission
is needed for meaningful comparison of different galaxies, 
a suitable alternative is to normalize $\FIivAfe$ by the generally larger and better 
constrained $24\mum$ flux or, better yet, the total infrared flux,

\begin{equation}\label{eq_tir}
\FTir = c \left( \zeta_1 \frac{\fMi}{\lambda_{\,24}} + 
   \zeta_2 \frac{\fMii}{\lambda_{\,70}} + 
   \zeta_3 \frac{\fMiii}{\lambda_{\,160}} \right).
\end{equation}

\noindent Here, $\FTir$ corresponds to the $3-1100\,\mum$ range,
$c$ is the speed of light,
$\lambda_{\,\mathrm{X}}$ refers to the effective wavelength of band X
($\lambda_{\,24} = 23.675\mum$, $\lambda_{\,70} = 71.440\mum$, 
$\lambda_{\,160} = 155.899\mum$), and the coefficients [$\zeta_1$,
$\zeta_2$, $\zeta_3$] are provided as a function of redshift 
($=[1.559, 0.7686, 1.347]$ for $z=0$) in Figure~7 of \citet{2002dal576apj159}.

\section{AFE in the Local Volume}\label{sec_lvlafe}

Using Equations~({\ref{eq_fIivAfe})$-$(\ref{eq_c3}) and the spectral 
modelling described in \S~\ref{sec_star}, the IRAC $8\mum$ photometry for 
the LVL galaxies was divided into contributions from aromatic features, 
the dust continuum, and starlight.  
Then, Equation~(\ref{eq_FIivAfe_final}) was used to determine the 
integrated flux in the five aromatic features 
(centered at 
$\lambda_{\mathrm{rest}} = 7.42, 7.60, 7.85, 8.33,\,\mathrm{and}\,8.61\,\mum$)
comprising the $8\mum$ complex, and the same systematic uncertainty obtained 
from the spectroscopic sample ($0.06\,\FIiv$) was assumed.
The results are provided in Table~\ref{tab_photometric}, 
and Figure~\ref{fig_lvlsed1} depicts the stellar and dust continuum
components relative to the observed photometry.

The reliability of this photometric approach was demonstrated for the 
40 SINGS galaxies comprising the spectroscopic sample (see \S~\ref{sec_specvsphot}); 
however, many of the LVL galaxies are significantly
less luminous.  Whereas the IRAC $8\mum$ band luminosity range
of the former is $9.9 > \log(\LIiv/\Lsun) > 7.0$, the LVL galaxies extend
from $\log(\LIiv/\Lsun)=9.6$ down to $\log(\LIiv/\Lsun)<4.3$.  The 
low-luminosity nature of this dwarf galaxy dominated sample is, in fact, what necessitated
a photometric prescription for measuring AFE in the first place.
However, the decomposition of the emission in the $8\mum$ band primarily
depends on the relative strengths of the other IRAC and MIPS $24\mum$ bands
(i.e., the shape or color of the mid-infrared SED).  The SINGS spectra (recall 
Figure~\ref{fig_spec1}) span the range from dust dominated SEDs with strong 
aromatic features to stellar dominated SEDs with little or no aromatic emission.
More specifically, the following discussion demonstrates the comparable color
ranges of the LVL and spectroscopic samples.

The estimation of the dust continuum at $8\mum$ in Equation~(\ref{eq_FIivAfe_final})
depends on $(\fIiiNs+\fIiiiNs)/(\fMiNs)$, which spans
$0.00-0.98$ and $0.06-0.73$ in the LVL and spectroscopic samples.
Likewise, the range of the $\fIiv/\fMi$ color, corresponding to the correction term given by
Equation~(\ref{eq_c3}), for the two samples is $0.03-2.72$ and $0.19-2.60$, respectively 
(note that for small $\fIiv/\fMi$ this correction becomes negligible).
More generally, there is very good agreement between the mid-infrared starlight--to--dust 
($\fIi/\fMi$) distributions of the 
LVL ($0.02-11.81$) and spectroscopic ($0.03-9.12$) galaxies.
These color comparisons exclude one outlier in the LVL sample.  The corresponding values
for UGC~05442 (see Figure~\ref{fig_lvlsed1}) 
are $(\fIiiNs+\fIiiiNs)/(\fMiNs) = 2.43$, 
$\fIiv/\fMi = 12.00$, and $\fIi/\fMi = 19.58$; however, the spectroscopic sample includes
several galaxies similarly dominated by starlight in the mid-infrared (see
Figure~\ref{fig_spec1}; e.g., NGC~4125).

\subsection{An AFE Inventory}\label{sec_inventory}

As demonstrated in \S~\ref{sec_sum}, the uncertainties in the photometrically 
determined $\FIivAfe$ are sufficiently random to allow for an accurate
inventory of the aromatic emission in a sample as large as LVL.  Summing
$\LIivAfe$ for all 258 galaxies, the total luminosity from the $8\mum$ 
complex of aromatic features is $9.49\times10^{43}\,\mathrm{erg/s}$ or 
$2.47\times10^{10}\,\Lsun$.  This includes upper limits for 38
galaxies and the assumption that all of the $8\mum$ light is aromatic for 
the 15 galaxies without constraints on $\FIivAfe$ 
(see Table~\ref{tab_galaxies}).  However, excluding these 53 galaxies decreases
the total luminosity by only 0.05\%, which is less than the
1.9\% propagated systematic uncertainty.

Figure~\ref{fig_cumafe} depicts this 
inventory in the form of a cumulative histogram, where the 10 galaxies with the 
highest $\LIivAfe$ are labeled.  It is worth noting that these 10 galaxies are 
responsible for 70\% of the total aromatic luminosity in LVL.  Likewise, 24 galaxies
account for 90\%.
From linear fits to $\LIivAfe$ versus the absolute B magnitude provided in
\citet{2009dal703apj517} and the total infrared luminosity given by 
Equation~(\ref{eq_tir}), the $10^{\mathrm{th}}/24^{\mathrm{th}}$ galaxy corresponds to 
$\mathrm{M}_\mathrm{B}$ = -18.82/-18.22 $\pm 0.87$ and 
$\LTir = 10^{9.54/9.25 \pm 0.17}\,\Lsun$.  Here, the uncertainties are taken to be the 
rms of the scatter about the fits.

\begin{figure*}[b]
\includegraphics[width=7.1in]{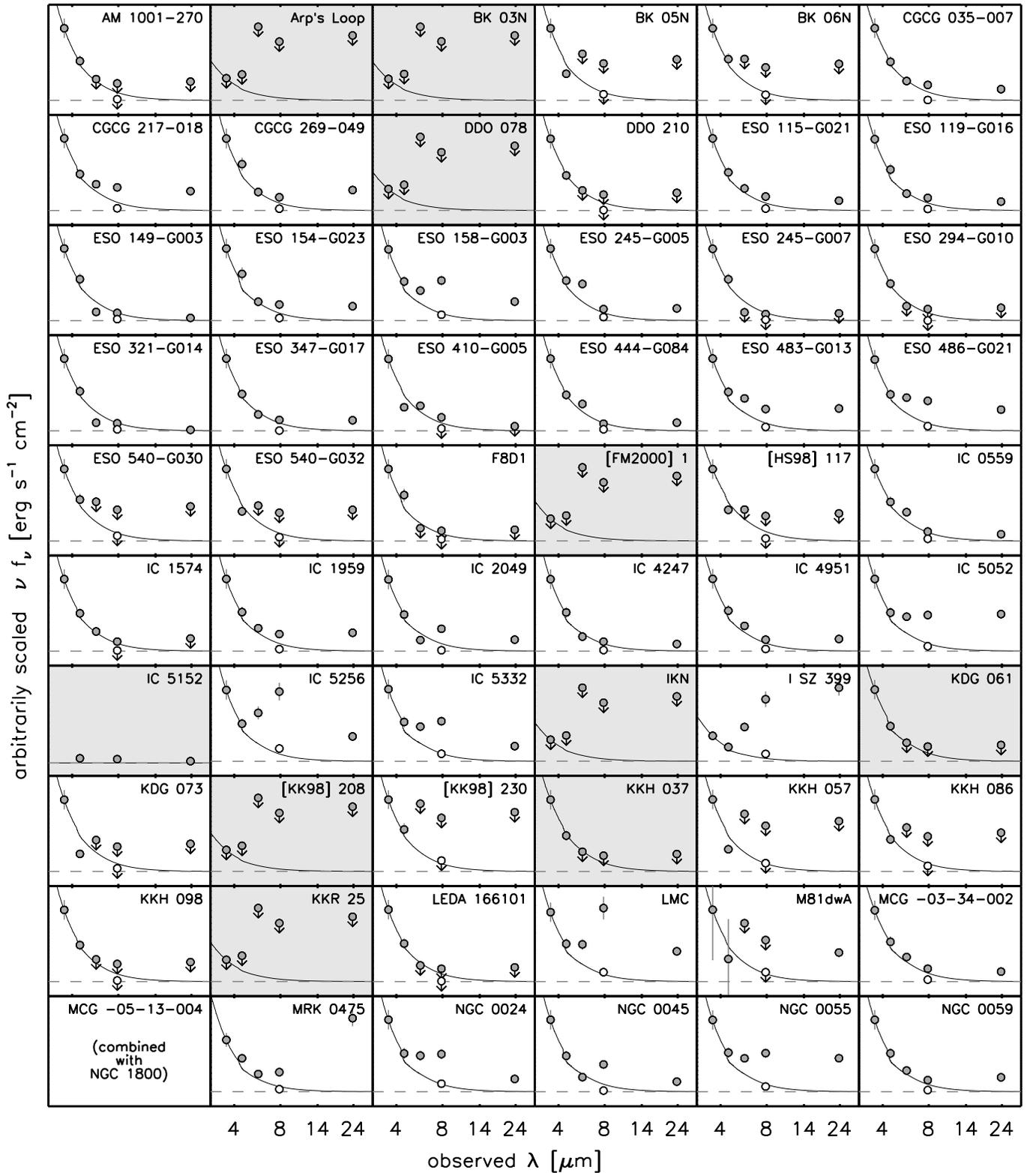}
\caption{Thumbnail SEDs for each LVL galaxy showing the IRAC 
  and MIPS $24\mum$ photometry (filled circles) with error bars, 
  the tailored stellar model (solid line), and the calculated dust continuum
  contribution to the IRAC $8\mum$ band (open circle).  Zero is 
  indicated by the dashed line, arrows pointing down from data 
  points signify upper limits, and shaded plots correspond to 
  those cases where the dust continuum contribution could not
  be determined.}
  \label{fig_lvlsed1}
\end{figure*}

\begin{figure*}[ht]
\setcounter{figure}{7}
\includegraphics[width=7.1in]{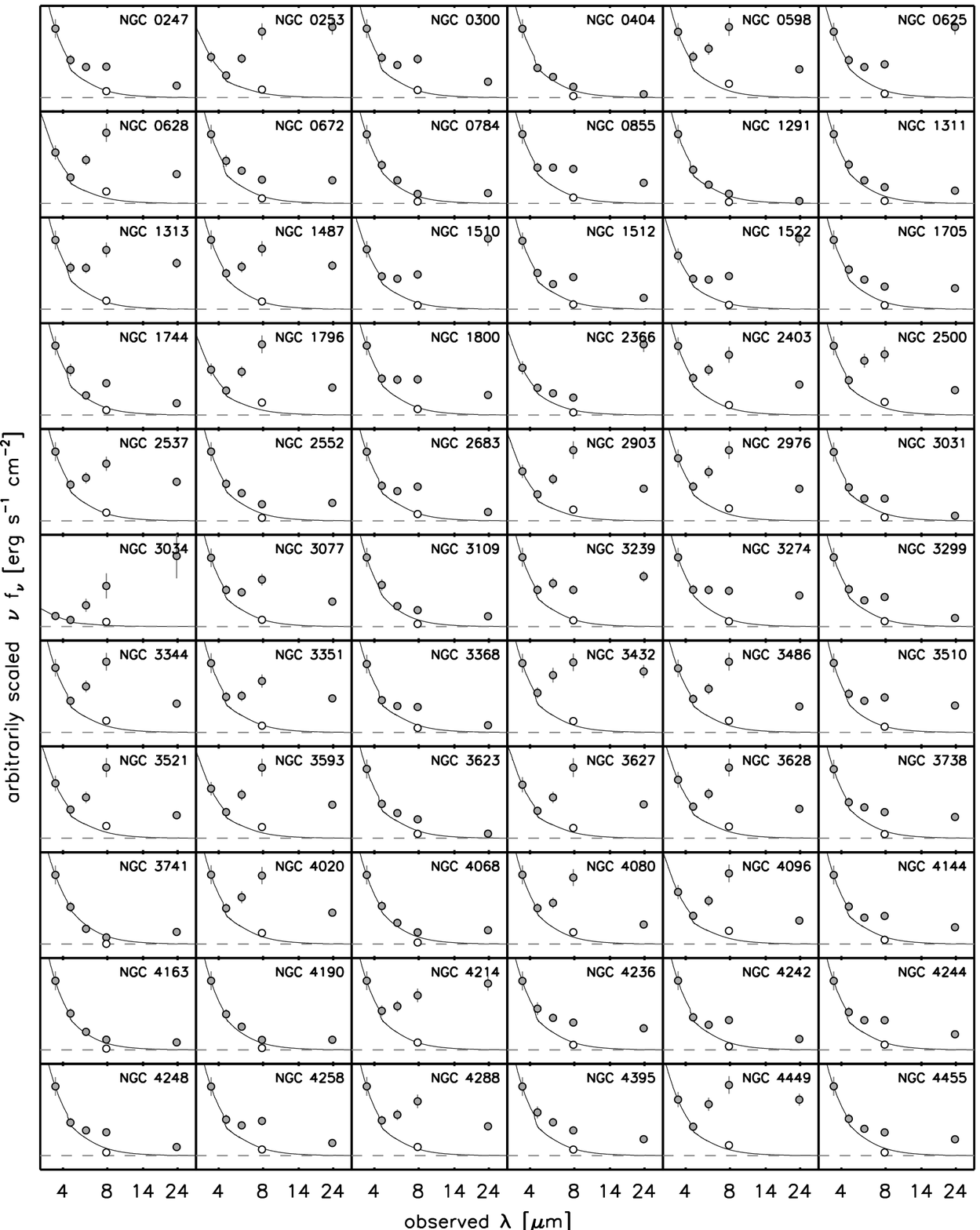}\\[5mm]
\centerline{FIG. \ref{fig_lvlsed1}.---\emph{Continued}}
\end{figure*}

\begin{figure*}[ht]
\setcounter{figure}{7}
\includegraphics[width=7.1in]{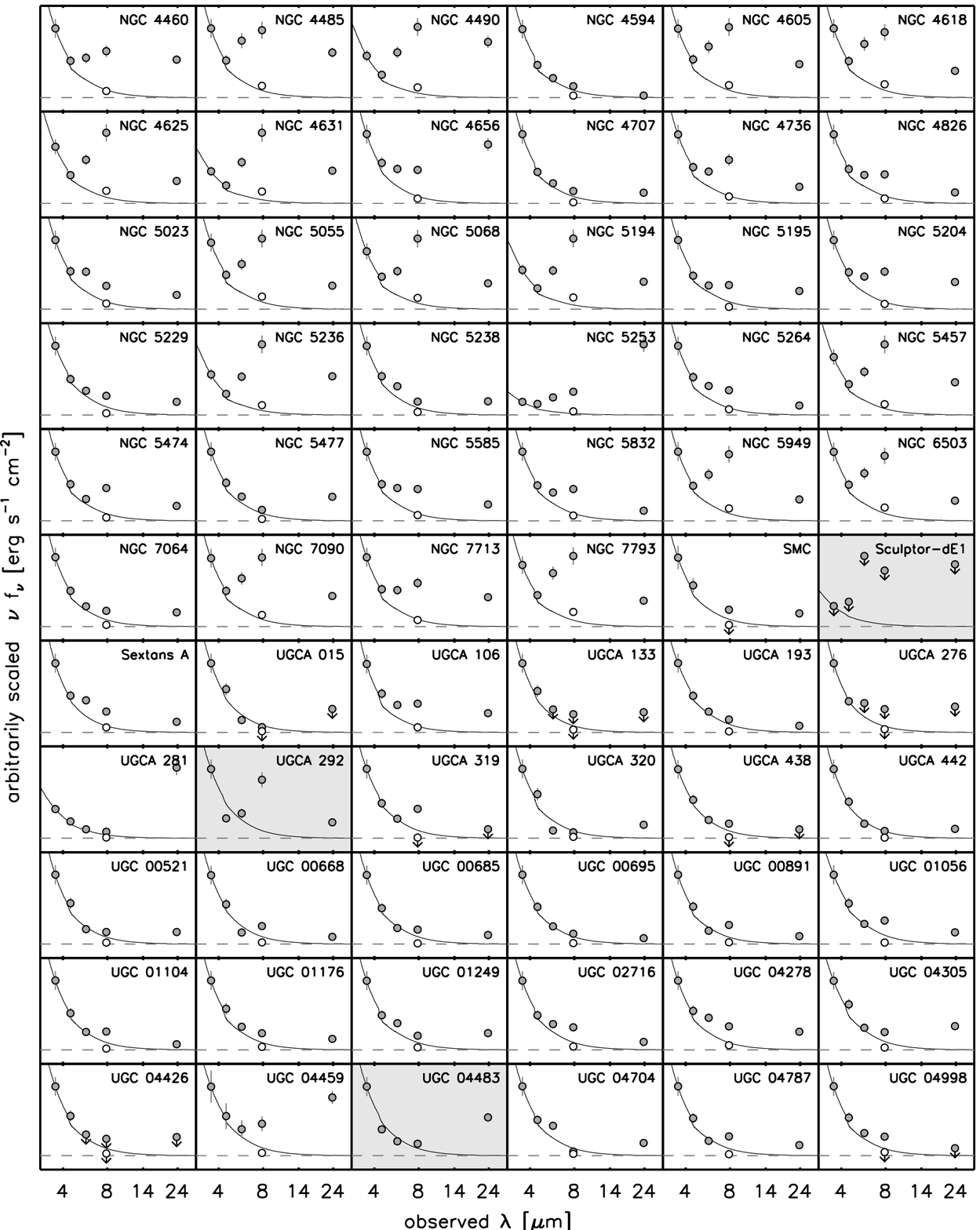}\\[5mm]
\centerline{FIG. \ref{fig_lvlsed1}.---\emph{Continued}}
\end{figure*}

\begin{figure*}[ht]
\setcounter{figure}{7}
\includegraphics[width=7.1in]{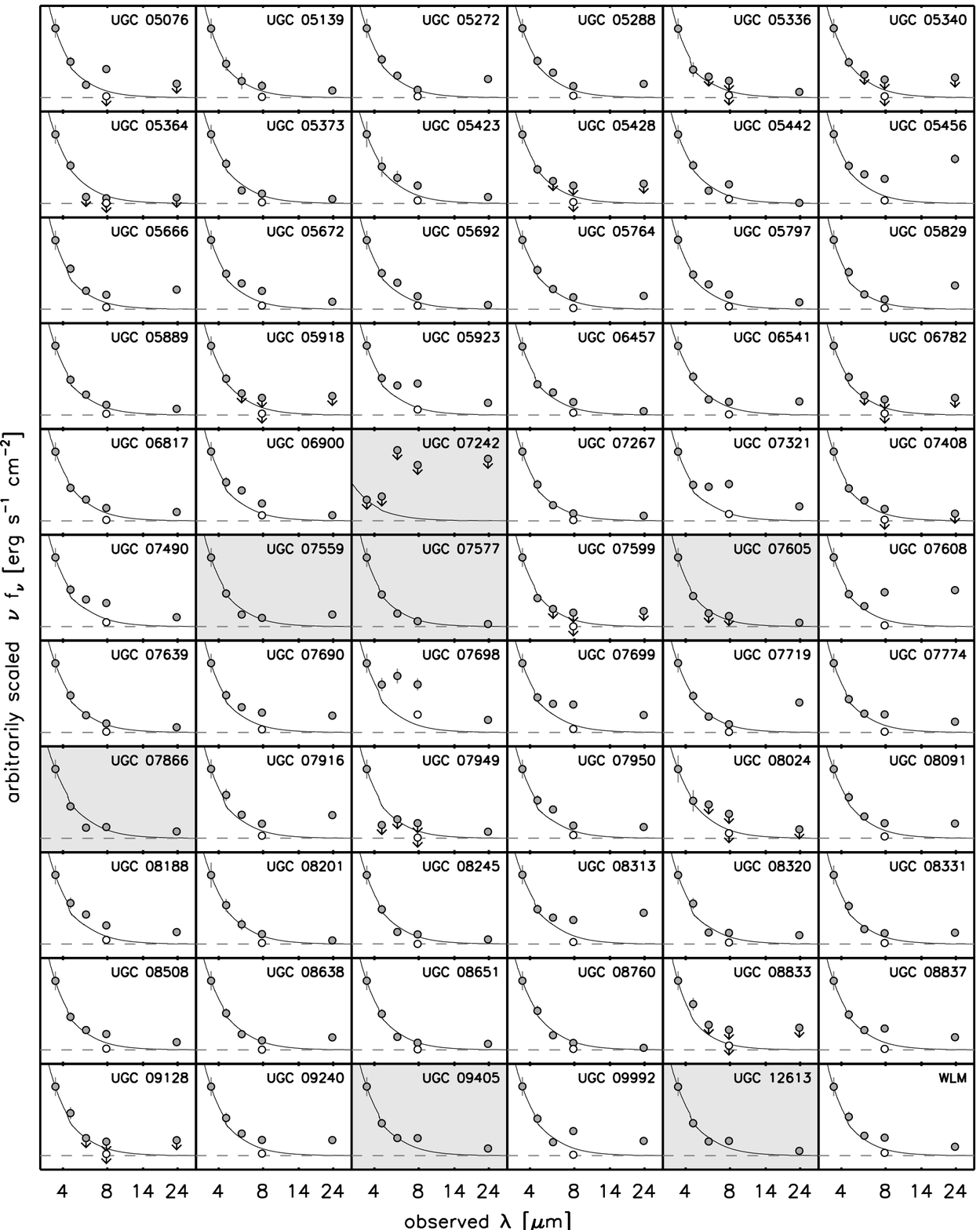}\\[5mm]
\centerline{FIG. \ref{fig_lvlsed1}.---\emph{Continued}}
\end{figure*}

\clearpage

\begin{figure}[t]
\setcounter{figure}{9}
\begin{center}
\includegraphics[width=3.3in]{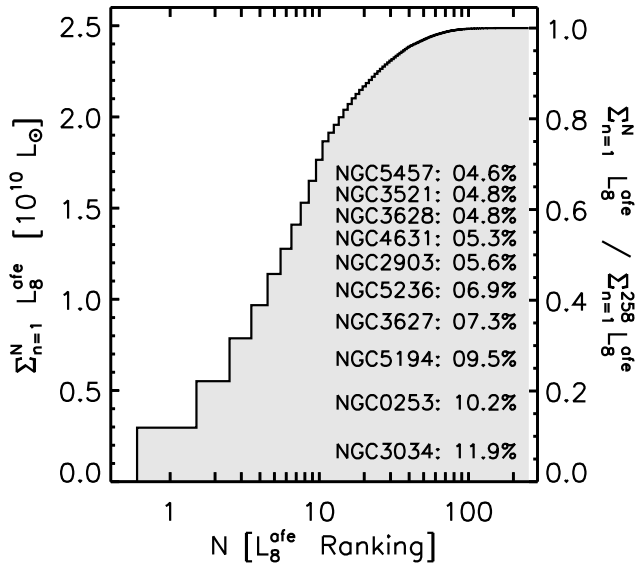}
\end{center}
\caption{Cumulative histogram of the photometrically determined 
  AFE luminosities for the 258 LVL galaxies (on absolute and relative
  scales on the left and right axes, respectively).  This is based on upper 
  limits for 43 galaxies and the assumption that $\FIivAfe\equiv\FIiv$
  for 10 galaxies without AFE measurements; however, the
  contributions of these 53 galaxies represent only 0.05\% of the total.
  The galaxies are ordered from left to right according to decreasing
  AFE luminosity, and the 10 highest contributors are labeled.}
  \label{fig_cumafe}
\end{figure}

\begin{figure*}[t] 
\begin{center}
\includegraphics[width=7.1in]{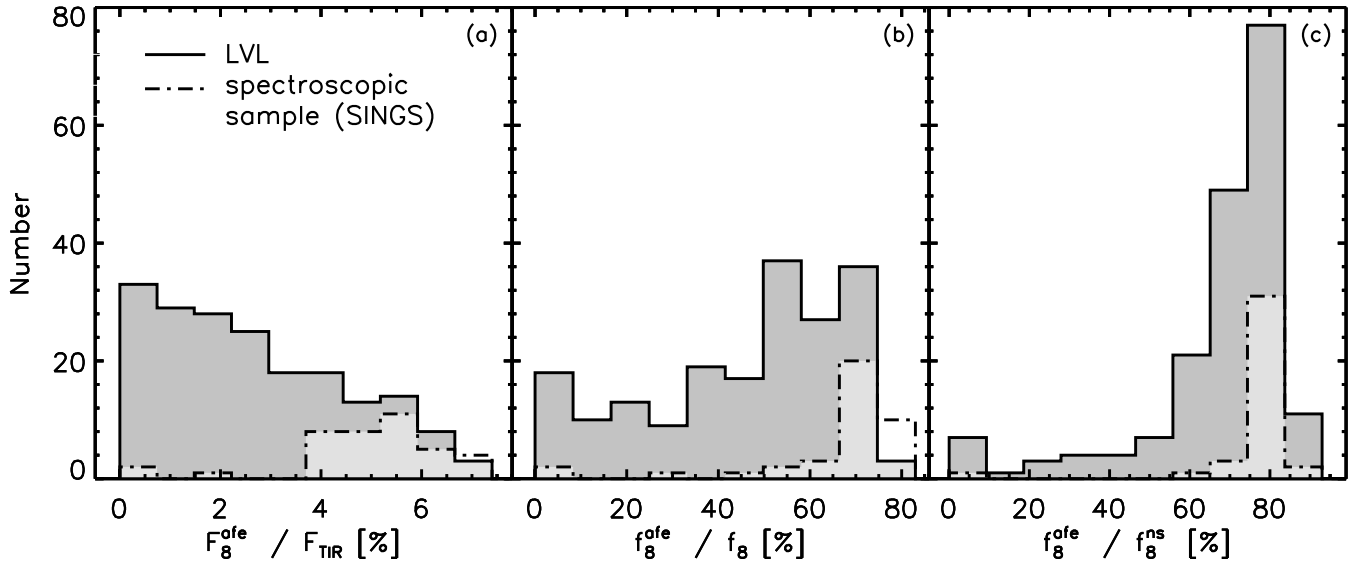}
\end{center}
\caption{Histograms of the photometrically determined $8\mum$ AFE
  relative to the 
  (a) total infrared ($3-1100\,\mum$) emission,
  (b) total IRAC $8\mum$ emission, and 
  (c) total IRAC $8\mum$ dust emission
  for the 40 SINGS galaxies in the spectroscopic sample and 
  the approximately 200 LVL galaxies with reliable values 
  (i.e., no upper limits for either quantity).}  
  \label{fig_afevstir}
\end{figure*}

The LVL survey represents 52\% of the local volume within 11 Mpc (see 
\S~\ref{sec_galaxies}), or approximately 2800 cubic Mpc.  Thus, the mean
density of aromatic emission from the $8\mum$ complex is 
$8.8\times10^6\,\Lsun\,\mathrm{Mpc}^{-3}$.  
Adopting the mean ratio $<\FIrsMirAfe/\FIrsIivAfe>\,=\,1.89$ from the 
spectroscopic sample (see Table~\ref{tab_spectroscopic}), the numbers 
above translate into a total LVL aromatic emission luminosity and mean 
density of $4.67\times10^{10}\,\Lsun$ and 
$1.7\times10^{7}\,\Lsun\,\mathrm{Mpc}^{-3}$, respectively,
for the mid-infrared wavelength range $5.5\mum<\lambda<20\mum$.
This scaling is based on the approximately inner kiloparsec of the
40 SINGS galaxies in the spectroscopic sample; however, the global photometry
used for the LVL galaxies is similarly dominated by the central region.

\subsection{Relative AFE}

The primary purpose of this paper is 
to demonstrate the efficacy of our photometric
prescription for measuring the strength of the $8\mum$ aromatic features and
to present such measurements for the LVL sample of galaxies.
However, we conclude this section by characterizing these LVL AFE measurements
relative to other galaxy properties addressed in this study.

\subsubsection{AFE Ratios}\label{sec_afe_fraction}

For the reasons outlined in \S~\ref{sec_ew}, casting the LVL aromatic feature
emission measurements as equivalent widths (i.e., dividing them by the
corresponding continua at $8\mum$) is not particularly illuminating.  Therefore, 
we consider instead their ratio to other less uncertain probes of dust emission:
the total infrared (see Equation~(\ref{eq_tir}) and \citet{2002dal576apj159}),
the total IRAC $8\mum$ emission, and the total dust emission at $8\mum$.

In comparison to the integrated luminosity in the aromatic features, the 
total $3-1100\,\mum$ infrared luminosity 
of the LVL galaxies is $\LTir = 4.86\times10^{11}\,\Lsun$.  This includes 59
upper limits which only amount to 0.12\%.
Thus, the fraction of the
total infrared luminosity in the local volume contributed by aromatic features
in the $8\mum$ complex and $5.5\mum<\lambda<20\mum$ range is
5.1 and 9.6 percent, respectively 
(recall the factor of 1.89 from \S~\ref{sec_inventory}).  
For an individual galaxy, the $8\mum$ complex
features contribute, on average, 2.5\% to the total infrared
(see Figure~\ref{fig_afevstir}a),  
approximately half that of the SINGS galaxies \citep{2007smi656apj770}. 

In contrast to $\FIivAfe/\FTir$, the ratio of $\fIivAfe$ over the total
IRAC $8\mum$ flux density shown in Figure~\ref{fig_afevstir}b skews towards 
larger values.  
This dominance of the aromatic features has resulted in the 
occasional use of the observed $8\mum$ emission as a proxy for AFE.
Such a generalization is clearly problematic given the substantial range in
$\fIivAfe/\fIiv$ from galaxy to galaxy.  For the LVL sample, this ratio
spans $0-80\,\%$ with an abrupt decline in values larger than 70\%.  The 
distribution becomes narrower when only non-stellar emission is considered
(see Figure~\ref{fig_afevstir}c); however, 
$\fIivAfe/\fIivNs \simeq \fIivAfe/(\fIivAfe+\fIivCnt)$ still 
potentially ranges from 0\% to greater than 90\%.
Note that the peak in the lowest bins of both Figures~\ref{fig_afevstir}b 
and \ref{fig_afevstir}c is likely the result of the approximately 6\% 
uncertainty coupled with our criterion that the measured
aromatic emission be greater than or equal to zero.

\subsubsection{AFE Ratio vs. TIR}\label{sec_afe_vs_tir}

The fact that the LVL galaxies have both a lower average luminosity and
a lesser mean $\FIivAfe/\FTir$ than the SINGS galaxies comprising the 
spectroscopic sample (Figure~\ref{fig_afevstir}a) suggests a relationship 
between these properties.  Figure~\ref{fig_afe_vs_tir} confirms that the 
aromatic feature emission to total infrared ratio is correlated 
with the total infrared luminosity
(the Spearman rank-order coefficient is $r_s = 0.66$),
albeit with significant scatter 
(despite the inverse dependence, the positive correlation with $\LTir$
is stronger than with the $24\mum$, $8\mum$, or B band 
luminosities).  
This dependence may (see \S~\ref{sec_afe_vs_oh})
simply be a repackaging of the luminosity-metallicity
relationship shown in Figure~\ref{fig_lvl_oh_vs_i2} and previously reported
trends between metallicity and the relative strength of the $8\mum$ 
aromatic features \citep[see, e.g.,][]{2008eng678apj804}.  However,
this result is unique with regard to the large sample size considered.
Note that the correlation weakens ($r_s = 0.54$) if the 
total dust at $8\mum$ is considered rather than just the aromatic portion.

\subsubsection{AFE Ratio vs. Oxygen Abundance}\label{sec_afe_vs_oh}

For a sample of starburst galaxies,
\citet{2008eng678apj804} found a weak correlation between oxygen abundance
and the equivalent width of emission from the same $8\mum$ complex
of aromatic features considered here.
A stronger correlation resulted when comparing
$\mathrm{EW}^{\,\mathrm{afe}}_{\,8}$ to the hardness of the radiation field
\citep[see, also,][]{2006jac646apj192}.  The opposite
was reported by \citet{2009wu0907arXiv1783} for the case of dwarf galaxies; 
however, their probe of aromatic emission was limited to the $\fIiv/\fIi$ 
color.  Looking at HII regions in M101, \citet{2008gor682apj336} found the 
same correlation with radiation field hardness as \citet{2008eng678apj804},
but no corresponding trend with metallicity.
 
For the LVL galaxies, we find a similar correlation as 
\citet{2008eng678apj804} between oxygen abundance and the relative strength 
of the AFE (Figure~\ref{fig_afe_vs_oh}).  There is significant
scatter; however, a Spearman rank-order test yields $r_s=0.71$.
Given the comparable trend with $\LTir$ found in
Figure~\ref{fig_afe_vs_tir} and the well-known L-Z
correlation shown in Figure~\ref{fig_lvl_oh_vs_i2}, this raises the question
of whether metallicity or luminosity is more fundamentally related to aromatic
feature emission.
The Spearman correlation coefficient is smaller for the total infrared
luminosity than the oxygen abundance, despite the approximately $0.1-0.2$ dex
systematic errors expected for the latter (see \S~\ref{sec_abun}), implying
that metallicity is the dominant factor. 
This is further supported by
partial correlation coefficients computd for the subset of galaxies with total 
infrared luminosity, aromatic emission, and oxygen abundance measurements.

\begin{figure}[b] 
\begin{center}
\includegraphics[width=3.3in]{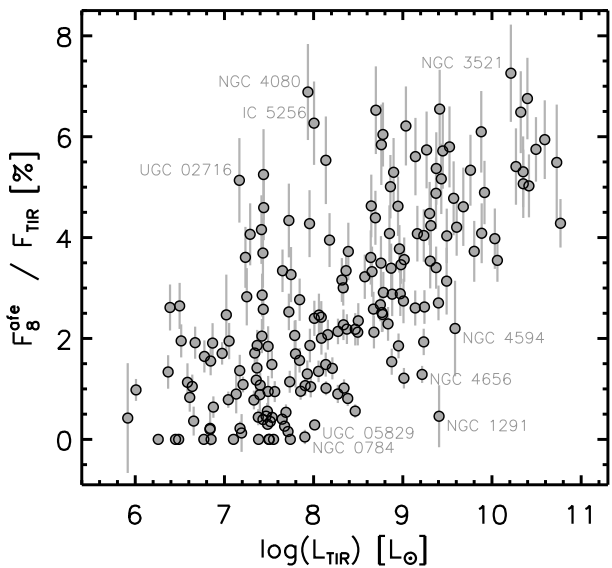}
\end{center}
\caption{
  The photometrically determined $8\mum$ AFE to
  total infrared ($3-1100\,\mum$) emission ratio
  versus the total infrared luminosity for the 189
  LVL galaxies unaffected by upper limits.  
  The corresponding Spearman rank-order correlation 
  coefficient is 0.66.}
  \label{fig_afe_vs_tir}
\end{figure}

The correlation with oxygen abundance is weaker when the ratio
of AFE to the total IRAC $8\mum$ flux is considered instead of $\FIivAfe/\FTir$
and is essentially gone when the stellar component of the IRAC $8\mum$ flux is
removed (Figure~\ref{fig_afefrac_vs_oh}).  
This may imply that while metallicity plays a role in the ratio of aromatic
molecules to total dust content (i.e., the TIR), 
the excitation of those molecules, and thus,
the relative amounts of AFE and dust emission at $8\mum$, are governed by other 
properties such as star formation and/or the hardness and intensity of the local 
radiation field.

\begin{figure}[b]
\begin{center}
\includegraphics[width=3.3in]{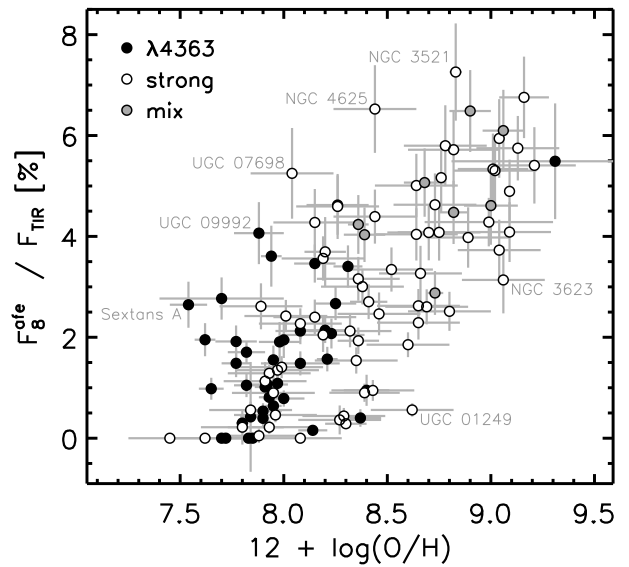}
\end{center}
\caption{
  The photometrically determined $8\mum$ AFE to
  total infrared ($3-1100\,\mum$) emission ratio
  versus oxygen abundance for the 114 
  LVL galaxies unaffected by upper limits or missing data. 
  Compiled from the literature, the oxygen abundances 
  (the shading of the circle reflects the 
  measurement method)
  have an expected systematic error of $0.1-0.2$
  in addition to the error bars shown here.
  The corresponding Spearman rank-order correlation 
  coefficient is 0.71.}
  \label{fig_afe_vs_oh}
\end{figure}

\begin{figure}[t]
\begin{center}
\includegraphics[width=3.3in]{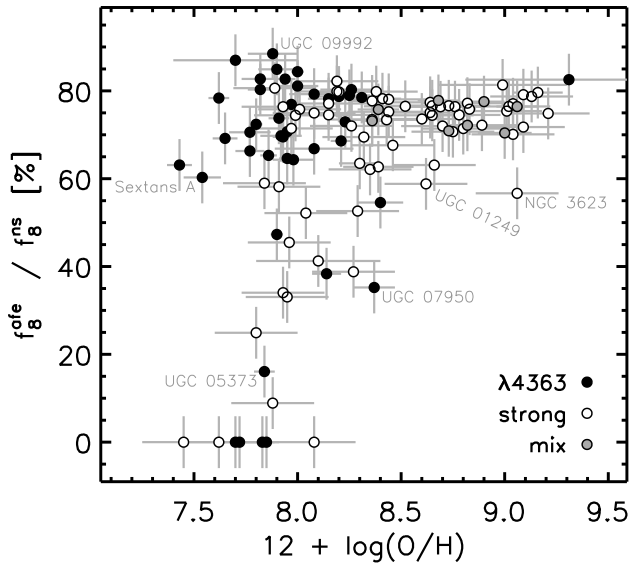}
\end{center}
\caption{
  The ratio of aromatic to total dust emission at $8\mum$
  versus oxygen abundance (the shading of the circle 
  reflects the measurement method).  Lack of a correlation
  (compared to Figure~\ref{fig_afe_vs_oh}) may imply that
  metallicity affects the abundance of aromatic molecules 
  relative to the total dust content, while other factors
  (such as star formation and/or the hardness of the radiation
  field) affect the excitation of the aromatic molecules.}
  \label{fig_afefrac_vs_oh}
\end{figure}

\section{Summary and Conclusions}\label{sec_concl}

We have presented a purely photometric methodology for measuring the strength
of the emission from mid-infrared aromatic features in nearby galaxies
that are either too faint, too numerous, or too spatially extended for
spectroscopy to be feasible.
\emph{Spitzer} IRAC (3.6, 4.5, and 5.8 $\mum$) and MIPS $24\mum$ photometry
are used to disentangle contributions to the IRAC $8\mum$ flux density from
starlight, the dust continuum, and the $8\mum$ complex of aromatic features.
The latter is then calibrated to approximate the integrated flux one would 
obtain spectroscopically for the
individual features at 7.42, 7.60, 7.85, 8.33, and 8.61 $\mum$.
With carefully
matched photometry, we demonstrate that the photometric technique recovers
the spectroscopically derived aromatic fraction of the total $8\mum$ luminosity
in a sample of 40 SINGS galaxies
with a maximal relative difference of 12\% and an rms of 6\%.  Given suitable 
5.5 and 10 $\mum$ photometry, the rms could decrease by a factor of two in
future studies.

The photometric approach was then used to measure aromatic emission 
from galaxies in the Local Volume Legacy sample, a \emph{Spitzer} IRAC and 
MIPS survey of a statistically complete and nearly volume-limited 
sample of 258 nearby 
galaxies within 11 Mpc.  
In doing so, we have presented the first inventory
of aromatic feature emission for a statistically complete sample of 
star-forming galaxies in the local universe.  
The total luminosity
in the LVL sample resulting from the $8\mum$ complex of five strong
aromatic features is 
$2.47\times10^{10}\,\Lsun$ with a mean volume density of 
$8.8\times10^6\,\Lsun\,\mathrm{Mpc}^{-3}$.
The corresponding 
values for all mid-infrared aromatic features in the wavelength range
$5.5\mum<\lambda<20\mum$ are larger by a factor of approximately 1.9.
Twenty-four of the LVL galaxies, corresponding to 
$\mathrm{M}_\mathrm{B} > -18.22 \pm 0.87$ and 
$\LTir > 10^{9.25 \pm 0.17}\,\Lsun$, are responsible for 90\% of the total
$8\mum$ aromatic luminosity.

For a given LVL galaxy, the $8\mum$ complex features 
contribute, on average, 2.5\% of the total infrared luminosity, or roughly 
half that of the more luminous SINGS galaxies.  As a whole, 
the combined contribution for the LVL sample is 5.1\%.  With respect to the 
total dust emission in the IRAC $8\mum$ band, the aromatic contribution
ranges from 0\% to 90\%.  Similarly, the aromatic features are responsible 
for between 0\% and 80\% of all the light in the IRAC $8\mum$ band.

Oxygen abundances compiled from the literature were presented for 129 of the 258
LVL galaxies.  These are shown to follow the well-known luminosity-metallicity
relationship with a best fit for the IRAC $4.5\mum$ band of
$12+\log\mathrm{(O/H)} = 5.06\pm0.04 - (0.164\pm0.002)\,\mathrm{M}_{4.5}$.
We find that the total infrared luminosity and oxygen abundance correlate 
weakly with the aromatic to total infrared flux ratio and tentatively identify 
the latter relationship as being more fundamental.  The trend with
oxygen abundance breaks down when the total dust emission at $8\mum$ is used to
scale the aromatic emission instead of the total infrared emission from 
$3-1100\,\mum$.  One possible explanation for this is that metallicity affects 
the 
abundance of aromatic molecules, while other factors such as star formation 
and/or the local radiation field determine their excitation.

\acknowledgements

This work is part of the \emph{Spitzer Space Telescope} Legacy Science Program
and was supported by National Aeronautics and Space Administration
(NASA) through contract 1336000 issued by the Jet Propulsion Laboratory (JPL),
California Institute of Technology (Caltech) under NASA contract 1407.
Additionally, this publication makes use of data
products from the Two-Micron All Sky Survey, which is a joint project of the 
University of Massachusetts and the Infrared Processing and Analysis 
Center (IPAC) at Caltech, funded by NASA and the National Science Foundation.
This research has made use of the NASA/IPAC Extragalactic Database, which is 
operated by JPL/Caltech, under contract with NASA.  Finally, we
thank the referee for helpful comments and suggestions 
that resulted in an improved paper.



\clearpage
{
  \LongTables
  \tabletypesize{\scriptsize}
\begin{deluxetable}{l ccc c c@{}r@{}cc@{}r@{}cc c ccc c cc}[t]
\tablecolumns{19}
\tablewidth{0pc}
\tablecaption{Galaxies.\label{tab_galaxies}}
\tablehead{
  \colhead{} &
  \multicolumn{3}{c}{Coordinates} &
  \colhead{} &
  \multicolumn{7}{c}{Distance} &
  \colhead{} &
  \colhead{} &
  \colhead{} &
  \colhead{} &
  \colhead{} &
  \colhead{} &
  \colhead{} \\
  \cline{2-4}\\[-0.25cm]
  \cline{6-12}\\[-0.25cm]
  \colhead{} &
  \colhead{$\alpha$} &
  \colhead{$\delta$} &
  \colhead{} &
  \colhead{} &
  \multicolumn{3}{c}{c$z$} &
  \multicolumn{3}{c}{D} &
  \colhead{} &
  \colhead{} &
  \multicolumn{3}{c}{Oxygen Abundance} &
  \colhead{} &
  \multicolumn{2}{c}{Subsample} \\ 
  \cline{14-16}\\[-0.25cm]
  \cline{18-19}\\[-0.25cm]
  \colhead{Galaxy} &
  \colhead{(J2000.0)} &
  \colhead{(J2000.0)} &
  \colhead{Ref.} &
  \colhead{} &
  \multicolumn{3}{c}{(km/s)} &
  \multicolumn{3}{c}{(Mpc)} &
  \colhead{Ref.} &
  \colhead{} &
  \colhead{12+Log$_{10}$(O/H)} &
  \colhead{Method\tablenotemark{a}} & 
  \colhead{Ref.} &
  \colhead{} &
  \colhead{LVL} &
  \colhead{IRS}
}
\startdata
WLM                  & 00:01:58.1 & $-$15:27:39 & 1 & & & -116 & & &    0.92 & & 1 & & $   7.77\pm0.10$ & $\lambda$4363 & 2$-$5 & & $\surd$ & \nodata \\
NGC~0024             & 00:09:56.7 & $-$24:57:44 & 1 & & & 554 & & &    8.13 & & 1 & & \nodata & \nodata & \nodata & & $\surd$ & \nodata \\
NGC~0045             & 00:14:04.0 & $-$23:10:55 & 1 & & & 471 & & &    7.07 & & 1 & & \nodata & \nodata & \nodata & & $\surd$ & \nodata \\
NGC~0055             & 00:14:54.0 & $-$39:11:49 & 1 & & & 129 & & &    2.17 & & 1 & & $   8.36\pm0.10$ & strong & 6$-$10 & & $\surd$ & \nodata \\
NGC~0059             & 00:15:25.4 & $-$21:26:42 & 1 & & & 382 & & &    5.30 & & 1 & & $   8.40\pm0.11$ & $\lambda$4363 & 11 & & $\surd$ & \nodata \\
ESO~410$-$G005       & 00:15:31.4 & $-$32:10:48 & 12 & & & \nodata & & &    1.90 & & 12 & & \nodata & \nodata & \nodata & & $\surd$ & \nodata \\
Sculptor$-$dE1       & 00:23:51.7 & $-$24:42:18 & 12 & & & \nodata & & &    4.20 & & 12 & & \nodata & \nodata & \nodata & & $\surd$ & \nodata \\
ESO~294$-$G010       & 00:26:33.3 & $-$41:51:20 & 12 & & & 117 & & &    1.90 & & 12 & & \nodata & \nodata & \nodata & & $\surd$ & \nodata \\
\enddata
\end{deluxetable}

}

\clearpage
\tabletypesize{\scriptsize}
\begin{turnpage}
  \begin{deluxetable}{l c c c c c c c c c c c c}
\tablecolumns{13}
\tablewidth{0pc}
\tablecaption{Spectroscopic Sample Data.\label{tab_spectroscopic}}
\tablehead{
  \colhead{} &
  \multicolumn{4}{c}{IRS-Aperture-Matched Photometry\tablenotemark{a}} &
  \colhead{} &
  \multicolumn{5}{c}{Synthetic Photometry} &
  \colhead{} &
  \colhead{} \\
  \cline{2-5}
  \cline{7-11}
  \colhead{} &
  \multicolumn{1}{c}{$\fIi$} &
  \multicolumn{1}{c}{$\fIii$} &
  \multicolumn{1}{c}{$\fIiii$} &
  \multicolumn{1}{c}{$\fIiv$} &
  \colhead{} &
  \colhead{$\fIrsMi$\tablenotemark{b}} &
  \multicolumn{1}{c}{$\fIivStr$\,\tablenotemark{c}} &
  \colhead{$\fIrsIivCnt$\,\tablenotemark{d}} &
  \colhead{$\fIrsIivAmo$\tablenotemark{d}} &
  \colhead{$\fIrsIivAfe$\tablenotemark{d,e}} &
  \colhead{$\FIrsIivAfe$\tablenotemark{d,e}} &
  \colhead{$\FIrsMirAfe$\,\tablenotemark{d}} \\
  \colhead{Galaxy} &
  \multicolumn{1}{c}{(mJy)} &
  \multicolumn{1}{c}{(mJy)} &
  \multicolumn{1}{c}{(mJy)} &
  \multicolumn{1}{c}{(mJy)} &
  \colhead{} &
  \colhead{(mJy)} &
  \multicolumn{1}{c}{(mJy)} &
  \colhead{(mJy)} &
  \colhead{(mJy)} &
  \colhead{(mJy)} &
  \colhead{($10^{-12} \mathrm{erg / s / cm}^2$)} &
  \colhead{($10^{-12} \mathrm{erg / s / cm}^2$)}
}
\startdata
MRK~0033 & 4.32E$+$1 & 3.19E$+$1 & 9.32E$+$1 & 2.48E$+$2 & &   1.29$\,\pm\,$0.02~E$+$3 & 1.08E$+$1 &  2.63$\,\pm\,$0.167~E$+$1 &   5.16$\,\pm\,$0.72~E$+$0 &  1.96$\,\pm\,$0.029~E$+$2 &  3.01$\,\pm\,$0.055~E$+$1 &  5.42$\,\pm\,$0.061~E$+$1 \\
NGC~0337 & 4.81E$+$1 & 3.37E$+$1 & 9.27E$+$1 & 2.26E$+$2 & &   4.24$\,\pm\,$0.07~E$+$2 & 1.17E$+$1 &  4.05$\,\pm\,$0.184~E$+$1 &   2.10$\,\pm\,$0.47~E$+$0 &  1.62$\,\pm\,$0.028~E$+$2 &  2.42$\,\pm\,$0.054~E$+$1 &  4.56$\,\pm\,$0.056~E$+$1 \\
NGC~0584 & 5.23E$+$2 & 3.13E$+$2 & 2.23E$+$2 & 1.36E$+$2 & &   5.87$\,\pm\,$0.10~E$+$1 & 1.31E$+$2 &  1.87$\,\pm\,$0.216~E$+$1 &   0.19$\,\pm\,$2.12~E$-$1 &  0.00$\,\pm\,$1.200~E$-$1 &  0.00$\,\pm\,$1.560~E$-$2 &  2.20$\,\pm\,$0.453~E$+$0 \\
NGC~0628 & 7.25E$+$1 & 4.70E$+$1 & 6.04E$+$1 & 1.15E$+$2 & &   7.38$\,\pm\,$0.12~E$+$1 & 1.72E$+$1 &  1.03$\,\pm\,$0.020~E$+$1 &   2.02$\,\pm\,$0.41~E$+$0 &  8.23$\,\pm\,$0.112~E$+$1 &  1.26$\,\pm\,$0.022~E$+$1 &  2.13$\,\pm\,$0.024~E$+$1 \\
NGC~1097 & 5.54E$+$2 & 3.88E$+$2 & 9.06E$+$2 & 2.27E$+$3 & &   5.30$\,\pm\,$0.08~E$+$3 & 1.38E$+$2 &  3.71$\,\pm\,$0.020~E$+$2 &   2.19$\,\pm\,$0.06~E$+$1 &  1.63$\,\pm\,$0.003~E$+$3 &  2.46$\,\pm\,$0.005~E$+$2 &  4.49$\,\pm\,$0.006~E$+$2 \\
NGC~1316 & 1.25E$+$3 & 7.70E$+$2 & 5.95E$+$2 & 4.17E$+$2 & &   3.86$\,\pm\,$0.06~E$+$2 & 3.08E$+$2 &  1.20$\,\pm\,$0.028~E$+$2 &   1.44$\,\pm\,$0.66~E$+$0 &  0.00$\,\pm\,$4.890~E$-$1 &  0.00$\,\pm\,$7.370~E$-$2 &  9.56$\,\pm\,$0.716~E$+$0 \\
NGC~1482 & 2.88E$+$2 & 2.15E$+$2 & 8.82E$+$2 & 2.47E$+$3 & &   4.83$\,\pm\,$0.08~E$+$3 & 7.71E$+$1 &  2.42$\,\pm\,$0.045~E$+$2 &   1.37$\,\pm\,$0.07~E$+$1 &  1.99$\,\pm\,$0.004~E$+$3 &  2.99$\,\pm\,$0.007~E$+$2 &  5.30$\,\pm\,$0.011~E$+$2 \\
NGC~1512 & 2.62E$+$2 & 1.66E$+$2 & 2.17E$+$2 & 4.07E$+$2 & &   4.68$\,\pm\,$0.07~E$+$2 & 6.43E$+$1 &  4.84$\,\pm\,$0.361~E$+$1 &   5.77$\,\pm\,$1.06~E$+$0 &  2.68$\,\pm\,$0.052~E$+$2 &  4.04$\,\pm\,$0.101~E$+$1 &  7.44$\,\pm\,$0.105~E$+$1 \\
\enddata
\tablecomments{
   The columns (left to right) are the 
   galaxy name, IRS aperture matched flux density 
   for the four IRAC channels, 
   synthetic MIPS $24\mum$ flux density obtained from the IRS 
   spectroscopy, synthetic IRAC $8\mum$ flux density for the 
   fitted starlight, dust continuum, atomic and molecular 
   line, and aromatic spectral contributions, 
   and integrated fluxes for the aromatic features 
   in both the $8\mum$ complex and the mid-infrared 
   ($5.5\mum < \lambda < 20\mum$).
   The compact format $X.XX\,\pm\,Y.YY\,\mathrm{E}Z$ 
   implies ($X.XX\,\pm\,Y.YY)\times10^{Z}$.
}
\tablenotetext{a}{One-sigma uncertainties of 10\% were adopted 
   for the IRS-aperture-matched photometry from comparison to synthetic 
   spectrophotometry at $24\mum$.}
\tablenotetext{b}{The uncertainty quoted for the synthetic MIPS $24\mum$ 
   photometry is the standard deviation for 1000 measurements based on random realizations of 
   the error in the IRS spectrum.}
\tablenotetext{c}{Because the stellar model is normalized at $3.6\mum$, 
   the uncertainty in $\fIivStr$ is the same 10\% described in footnote a.}
\tablenotetext{d}{The uncertainty represents the standard deviation in 
   measurements derived from independent PAHFIT results 
   for 1000 random realizations of the error in the IRS spectrum.}
\tablenotetext{e}{The five aromatic features at 7.42, 7.60, 7.85, 
   8.33, and 8.62 $\mum$ (that fall entirely within the IRAC $8\mum$ band) are included, but not 
   the two (partially overlapping) features at 6.22 and 6.69 $\mum$.}
\end{deluxetable}

\end{turnpage}

\clearpage
{
  \LongTables
  \tabletypesize{\scriptsize}
\begin{deluxetable}{l c c c c c c c c c}
\tablecolumns{10}
\tabletypesize{\tiny}
\tablewidth{0pc}
\tablecaption{LVL Photometric Data.\label{tab_photometric}}
\tablehead{
  \colhead{} &
  \colhead{} &
  \multicolumn{4}{c}{Synthetic Stellar Photometry\tablenotemark{b}} & 
  \colhead{} &
  \colhead{} &
  \colhead{} &
  \colhead{} \\
  \cline{3-6}\\[-0.25cm]
  \colhead{} &
  \colhead{$\fIiv$\tablenotemark{a}} & 
  \colhead{$\fIiiStr$} &
  \colhead{$\fIiiiStr$} &
  \colhead{$\fIivStr$} &
  \colhead{$\fMiStr$} &
  \colhead{$\fIivCnt$} &
  \colhead{$\fIivAfe$} &
  \colhead{$\FIivAfe$} &
  \colhead{$\FTir$} \\
  \colhead{Galaxy} &
  \colhead{(mJy)} &
  \colhead{(mJy)} &
  \colhead{(mJy)} &
  \colhead{(mJy)} &
  \colhead{(mJy)} &
  \colhead{(mJy)} &
  \colhead{(mJy)} &
  \colhead{($10^{-12} \mathrm{erg / s / cm}^2$)} &
  \colhead{($10^{-12} \mathrm{erg / s / cm}^2$)}
}
\startdata
NGC~3077             & 8.13$\,\pm\,$1.01~E$+$2 & 3.18E$+$2 & 2.13E$+$2 & 1.29$\,\pm\,$0.17~E$+$2 & 1.48E$+$1 & 1.18$\,\pm\,$0.53~E$+$2 & 5.30$\,\pm\,$0.49~E$+$2 & 8.11$\,\pm\,$0.75~E$+$1 & 1.62$\,\pm\,$0.14~E$+$3 \\
UGC~05336            & $<$4.00E$+$0 & 3.79E$+$0 & 2.74E$+$0 & 1.63$\,\pm\,$0.24~E$+$0 & 1.80E$-$1 & $<$5.38E$-$1 & $<$1.72E$+$0 & $<$2.63E$-$1 & 7.82$\,\pm\,$1.17~E$+$0 \\
UGC~05288            & 3.14$\,\pm\,$0.40~E$+$0 & 5.37E$+$0 & 3.58E$+$0 & 2.11$\,\pm\,$0.29~E$+$0 & 2.49E$-$1 & 5.08$\,\pm\,$2.04~E$-$1 & 4.90$\,\pm\,$1.88~E$-$1 & 7.49$\,\pm\,$2.89~E$-$2 & 1.90$\,\pm\,$0.20~E$+$1 \\
AM~1001$-$270        & $<$1.70E$+$0 & 2.12E$+$0 & 1.41E$+$0 & 8.20$\,\pm\,$1.11~E$-$1 & 9.91E$-$2 & $<$1.16E$-$1 & $<$7.16E$-$1 & $<$1.09E$-$1 & $<$1.10E$+$1 \\
NGC~3274             & 2.49$\,\pm\,$0.31~E$+$1 & 1.36E$+$1 & 9.05E$+$0 & 5.36$\,\pm\,$0.73~E$+$0 & 6.29E$-$1 & 3.71$\,\pm\,$1.62~E$+$0 & 1.48$\,\pm\,$0.15~E$+$1 & 2.27$\,\pm\,$0.23~E$+$0 & 9.37$\,\pm\,$0.82~E$+$1 \\
UGC~05076            & 4.57$\,\pm\,$0.60~E$+$0 & 3.00E$+$0 & 1.49E$+$0 & 1.11$\,\pm\,$0.15~E$+$0 & 1.26E$-$1 & $<$2.06E$-$1 & $<$3.05E$+$0 & $<$4.66E$-$1 & $<$1.30E$+$1 \\
NGC~5023             & 3.09$\,\pm\,$0.39~E$+$1 & 2.54E$+$1 & 1.71E$+$1 & 1.02$\,\pm\,$0.14~E$+$1 & 1.19E$+$0 & 7.52$\,\pm\,$2.01~E$+$0 & 1.23$\,\pm\,$0.19~E$+$1 & 1.89$\,\pm\,$0.28~E$+$0 & 1.01$\,\pm\,$0.10~E$+$2 \\
NGC~4214             & 5.47$\,\pm\,$0.68~E$+$2 & 1.90E$+$2 & 1.27E$+$2 & 7.57$\,\pm\,$1.02~E$+$1 & 8.73E$+$0 & 7.34$\,\pm\,$3.56~E$+$1 & 3.73$\,\pm\,$0.33~E$+$2 & 5.71$\,\pm\,$0.50~E$+$1 & 2.14$\,\pm\,$0.19~E$+$3 \\
\enddata
\tablecomments{
   The columns (left to right) are the 
   galaxy name, IRAC $8\mum$ flux density, modelled synthetic  
   IRAC 4.5, 5.8 and 8 $\mum$ and MIPS $24\mum$ stellar 
   photometry, computed dust continuum and $8\mum$ aromatic feature complex
   contributions to the IRAC $8\mum$ band, integrated 
   aromatic emission for the $8\mum$ complex, and 
   the total infrared flux between $3\mum$ and $1100\mum$.
   The compact format $X.XX\,\pm\,Y.YY\,\mathrm{E}Z$ 
   implies ($X.XX\,\pm\,Y.YY)\times10^{Z}$.
}
\tablenotetext{a}{Photometry for the other \emph{Spitzer} bands 
   can be found in the LVL survey description and infrared photometry 
   paper \citep{2009dal703apj517}.}
\tablenotetext{b}{The uncertainties in the synthetic stellar photometry 
  are the propogated uncertainties in the IRAC $3.6\mum$ photometry, thus 
  the uncertainty in $\mathrm{f}^{\,\mathrm{str}}_{\,\mathrm{X}}$ is simply the provided uncertainty in $\fIivStr$ 
  multiplied by the ratio of $\mathrm{f}^{\,\mathrm{str}}_{\,\mathrm{X}} / \fIivStr$.}
\tablenotetext{c}{The NGC~1800 entry is the combined photometry 
  of that galaxy and MCG~-05-13-004 (which spatially overlap).}
\end{deluxetable}

}

\end{document}